\documentclass{aa}  

\usepackage{graphicx}
\usepackage{txfonts}
\usepackage{fontawesome}
\usepackage{amsmath}
\usepackage{amssymb}
\usepackage{mathtools}
\usepackage{microtype}
\usepackage{graphicx}
\usepackage{subfigure}
\usepackage{caption}
\usepackage{booktabs} 
\usepackage{fontawesome}
\usepackage{hyperref}
\usepackage{xcolor}
\usepackage{multirow}
\usepackage{bigdelim}
\usepackage{float}
\begin{document}

   \title{Ground-based image deconvolution with Swin Transformer UNet}


   \author{U. Akhaury
          \inst{1}
          \and
          P. Jablonka
          \inst{1}
          \and
          J.-L. Starck
          \inst{2,3}
          \and
          F. Courbin
          \inst{1}
          }

   \institute{Laboratory of Astrophysics, Ecole Polytechnique Fédérale de Lausanne (EPFL), Observatoire de Sauverny, CH-1290 Versoix, Switzerland.
              \email{utsav.akhaury@epfl.ch}
         \and
             Université Paris-Saclay, Université Paris Cité, CEA, CNRS, AIM, 91191, Gif-sur-Yvette, France
             \and
             Institutes of Computer Science and Astrophysics, Foundation for Research and Technology Hellas (FORTH), Greece \\
             }


 
  \abstract
   {}
   {As ground-based all-sky astronomical surveys will gather millions of images in the coming years, a critical requirement emerges for the development of fast deconvolution algorithms capable of efficiently improving the spatial resolution of these images. By successfully recovering clean and high-resolution images from these surveys, the objective is to deepen the understanding of galaxy formation and evolution through accurate photometric measurements.}
   {We introduce a two-step deconvolution framework using a Swin Transformer architecture. Our study reveals that the deep learning-based solution introduces a bias, constraining the scope of scientific analysis. To address this limitation, we propose a novel third step relying on the active coefficients in the sparsity wavelet framework.}
   {We conducted a performance comparison between our deep learning-based method and Firedec, a classical deconvolution algorithm, based on an analysis of a subset of the EDisCS cluster samples. We demonstrate the advantage of our method in terms of resolution recovery, generalisation to different noise properties, and computational efficiency. The analysis of this cluster sample not only allowed us to assess the efficiency of our method, but it also enabled us to quantify the number of clumps within these galaxies in relation to their disc colour. This robust technique that we propose holds promise for identifying structures in the distant universe through ground-based images.}
   {}


   \maketitle
%

\section{Introduction}

High spatial resolution and high signal-to-noise observations are prerequisites to most observational astrophysical problems. However, expecting the two conditions to happen simultaneously is challenging, as space telescopes have a limited collecting power, while large telescopes are ground-based and are therefore affected by atmospheric turbulence. This is clearly illustrated by the two main missions of the next decade: the ESA-NASA \textit{Euclid} space telescope \citep{Euclid2, Euclid1} and the \textit{Vera C. Rubin} Observatory \citep{LSST}. Exploiting the best of both worlds is possible, provided reliable post-processing techniques are developed to remove blurring by the atmosphere and instrument point spread function (PSF). To further complicate matters, sensor variations introduce noise into images. Hence, image deconvolution in astrophysics is an ill-posed and ill-conditioned inverse problem that requires regularisation to achieve a unique solution. This was realised very early in the field, and solutions were proposed, such as minimising the Tikhonov function \citep[e.g.][]{tikhonov1977solutions} or maximizing the entropy of the solution \citep[e.g.][]{skilling1984}. Other algorithms, based on Bayesian statistics, include the Lucy-Richardson algorithm, used on the early data from the \textit{Hubble} Space Telescope (HST; \cite{Richardson1972}; \cite{Lucy1974}). \citet[MCS;][]{MCS} proposed a two-channel method that separates the point sources from the spatially extended ones and deconvolves using a narrow PSF, hence achieving finite but improved resolution compatible with the sampling (pixel) chosen to represent the solution. Improvements of the `MCS' method implement wavelet regularisation of the extended channel of the solution \cite{firedec}, and this improved method was further refined by \citet[STARRED;][]{STARRED}, who used an isotropic wavelet basis called Starlets \citep{Starck2015} to regularise the solution.

However, deep learning offers a completely different approach, by learning the properties of the desired solution. Once trained, deep learning-based methods are also orders of magnitude faster than classical methods. Of note, UNets \citep{ronnenberger2015Unet} have become popular due to their highly non-linear processing and multi-scale approach. Building on Unet, \cite{sureau2020} developed the Tikhonet method for deconvolving galaxy images in the optical domain. They demonstrated that Tikhonet outperformed sparse regularisation-based methods in terms of mean squared error and a shape criterion, where a measure of the galaxy ellipticity was used to encode its shape. \citet{shapenet} added a shape constraint to the Tikhonet loss function and achieved better performance. In our recent work \citep{akhaury2022}, we have proposed a new deconvolution approach that employs the Learnlet decomposition \citep{ramzi:hal-03346892}. It uses the same two-step approach as in \citet{sureau2020} but substitutes the UNet denoiser by Learnlet. 

With the recent advent of Vision Transformers \citep{vit}, significant progress has been made in the field of image restoration \citep{swinir, restormer, uformer}. This is the motivation for the present work to investigate the performance of SUNet \citep{sunet}, a variant of Unet with Swin Transformer blocks \citep{swintrans} replacing the convolutional layers. To our knowledge, SUNet has not yet been used as a denoiser in a deconvolution framework. We show that the neural network outputs introduce bias, thereby limiting the scientific analysis. The bias appears as a small flux loss in the outputs. To counter this, we propose a third debiasing procedure based on the active coefficients in the sparsity wavelet framework called multi-resolution support \citep{MRSupp}, explained in Section \ref{MRS}. Our experiments involved real HST images, and the network was trained on images extracted from the CANDELS survey \citep{CANDELS, CANDELS_HST}.

Finally, to assess its generalisability, we tested our new code on a completely different sample of ground-based images obtained with the FORS2 camera at the Very Large Telescope (VLT). Spatially resolved images of galaxies can help address a multitude of topics involving their morphologies. In the present case, we tackle the question of how the galaxy cluster environment can impact the properties of the discs of spiral galaxies.  From the same dataset, \cite{cantale} found that $50\%$ of cluster spiral galaxies at redshift $0.5$-$0.9$ have disc $V-I$ colours that are redder by more than $1\sigma$ of the mean colours of their field counterparts. This result was obtained thanks to the deconvolution code Firedec \citep{firedec}. The VLT $V$- and $I$-band images, with initial spatial resolutions between $0.4\arcsec$ and $0.8\arcsec$, were deconvolved with a target final resolution of $0.1\arcsec$ on $0.05\arcsec$ pixels. Even though the gain in resolution was indeed substantial, conditions were not sufficient to go beyond the global photometric properties of the discs. In this study, we go one step further and investigate their internal structure, specifically by identifying their star-forming clumps.


Compact star-forming clumps have been identified in distant galaxies, particularly with the aid of HST deep images \citep[e.g.,][]{Wuyts2012, Guo2015, Sattari2023}. They are understood to play a crucial role in galaxy assembly and star-formation activity. Recent research by \cite{Sok_2022} explored clump fractions in star-forming galaxies from multi-band analysis in the COSMOS field. The match between HST and ground-based resolution was performed with Firedec. Their findings indicated a decline in the fraction of clumpy galaxies with increasing stellar masses and redshifts. Moreover, they observed that clumps contributed a higher fractional mass towards galaxies at higher redshifts. In our study, by employing a more powerful deconvolution algorithm capable of accurately recovering small-scale structures at high spatial resolution from ground-based multi-band observations (as demonstrated in Section \ref{sec:dec_comparison}), our goal is to quantify the number of clumps in EDisCS cluster galaxies and examine their relationship with disc colour. 
 
In Section \ref{dec_problem}, we present the deconvolution problem and introduce our proposed deep learning method to address it. The process of generating our dataset and conducting experiments is outlined in Section \ref{sec:data_exp}. In Section \ref{results}, we present the results of our deconvolution algorithm.
Finally, we draw our conclusions in Section \ref{concl}. To support reproducible research, the codes and trained models utilised in this article are publicly accessible in Section \ref{sec:reproducible_research}. Additional studies and supplementary information can be found in Appendices \ref{sec:supp_fig}, \ref{sec:MRS_appendix}, and \ref{sec:hallucintaions}.

\section{Deep learning-based deconvolution}
\label{dec_problem}

The deconvolution problem can be summarised with a very simple (but hard to solve) equation. Let $\mathbf{y}\in\mathbb{R}^{n\times n}$ be the observed image and $\mathbf{h}\in\mathbb{R}^{n\times n}$ be the PSF. The observed image can be modelled as
\begin{equation}
 \label{eq:invprob}
  \mathbf{y} = \mathbf{h} \ast \mathbf{x_t} + \mathbf{\eta},
\end{equation}
where $\mathbf{x_t} \in \mathbb{R}^{n \times n}$ denotes the target image, $\mathbf{*}$ denotes the convolution operation, and $\mathbf{\eta} \in\mathbb{R}^{n\times n}$ denotes additive Gaussian noise. The goal is to recover the ground truth image $\mathbf{x_t}$, given the PSF convolution and the unknown noise. Such ill-posed inverse problems require regularisation of the solution in order to select the one that is most appropriate compared to the many others that are compatible with the data. Sparse wavelet regularisation using the $\ell_0$ or $\ell_1$ norm remained the most accepted regularisation in the past, but the recent advent of machine learning methods has changed the paradigm. 

\subsection{Tikhonov deconvolution}
\label{sec:tikho}
Tikhonov Deconvolution is a two-step deep learning-based deconvolution technique. In the first step, the input images undergo deconvolution using a Tikhonov filter with quadratic regularisation. If $\mathbf{H}\in\mathbb{R}^{n^2 \times n^2}$ denotes the circulant matrix associated with the convolution operator $\mathbf{h}$, the Tikhonov solution of equation \ref{eq:invprob} is expressed as
\begin{equation}
    \label{eq:tikhonov}
    \mathbf{\hat{x} = \left(H^\top H+\lambda \Gamma^\top \Gamma\right)^{-1}H^\top y\quad}.
\end{equation}
Here, $\mathbf{\Gamma}\in\mathbb{R}^{n^2 \times n^2}$ represents the linear Tikhonov filter, configured as a Laplacian high-pass filter to penalise high frequencies. The regularisation weight, denoted as $\mathbf{\lambda}\in\mathbb{R}_+$, is determined through a grid search. The Tikhonov step produces deconvolved images containing correlated additive noise, which is subsequently eliminated in the following step by an appropriate denoiser. The denoisers are trained to learn the mapping from the Tikhonov output $\hat{\mathbf{x}}$ to the ground truth image $\mathbf{x_t}$ by minimising a suitable loss function, such as $\ell_1$ or $\ell_2$.

The denoising performance is significantly influenced by the choice of the model architecture. To effectively capture distant correlations, it is crucial to incorporate multi-scale processing in the model design. This consideration leads to the adoption of a layout similar to that of a UNet \citep{ronnenberger2015Unet}. Additionally, \citet{Mohan2020Robust} demonstrated that biases in convolutional layers can result in a low generalisation capability. Consequently, for our experiments, we opted for bias-free networks.

\subsection{SUNet denoising}

\begin{figure}[h!]
\begin{center}
\includegraphics[width=\columnwidth]{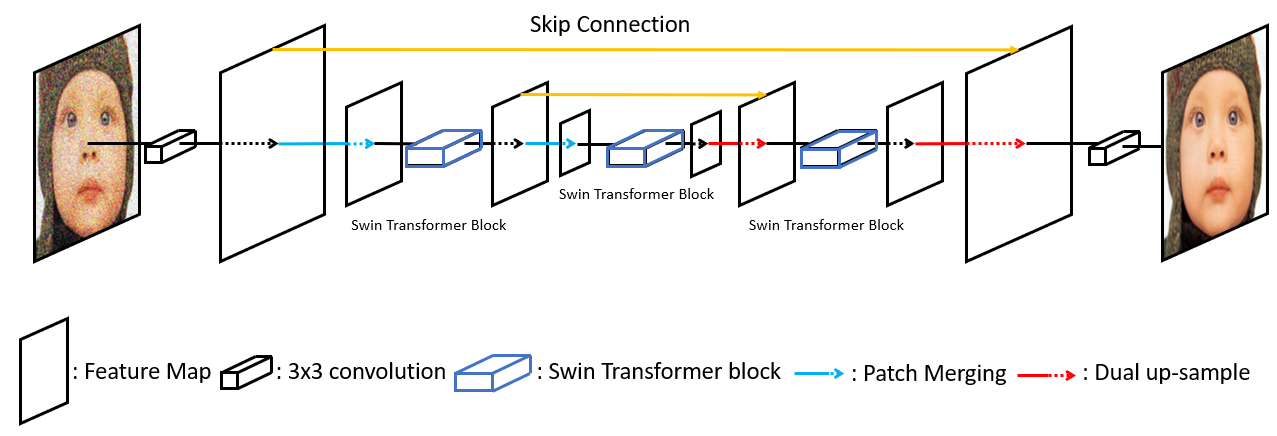}
\end{center}
\caption{\label{fig:sunet_arch} {SUNet architecture with Swin Transformer blocks replacing the convolutional layers while preserving the multi-scale Unet backbone. Credits: \citet{sunet}}}
\end{figure}

In recent times, the UNet architecture has become popular in various image-processing applications due to its incorporation of hierarchical feature maps, which facilitate the acquisition of multi-scale contextual features. It is widely employed in diverse computer vision tasks, including segmentation and restoration \citep{yu2019deep, gurrola2021unet}. Evolved versions such as Dense-UNet \citep{fdunet}, Res-UNet \citep{Xiao2018WeightedRF}, Non-local UNet \citep{deephdr}, and Attention UNet \citep{raunet} have also emerged. Thanks to its flexible structure, UNet can adapt to different building blocks, enhancing its overall performance. Moreover, the evolution of image-processing methodologies has seen the introduction of transformer models \citep{transformer}, which were initially successful in natural language processing but have also demonstrated impressive performance in image classification \citep{vit, yuan2021tokenstotoken}. However, when directly applied to vision tasks, transformers face challenges such as the large-scale difference between images and sequences, making them less effective in modelling long sequences due to a need for a square number of parameters for one-dimensional sequences. Additionally, transformers are not well suited for dense prediction tasks such as instance segmentation at the pixel level. Swin Transformer addresses these issues by introducing a shifted-window mechanism to reduce the number of parameters, establishing itself as a state-of-the-art solution for high-level vision tasks \citep{swintrans}. Taking inspiration from these advancements, \citet{sunet} incorporated Swin Transformers as building blocks within the UNet architecture and showed that they could achieve competitive results compared to existing benchmarks for image denoising. The network architecture is visually depicted in Figure \ref{fig:sunet_arch}, and the PyTorch code is publicly available on GitHub (details in Section \ref{sec:reproducible_research}).

While SUNet was originally developed for white Gaussian noise removal, the Tikhonov deconvolution step (equation \ref{eq:tikhonov}) introduces alterations to the image's noise characteristics. Consequently, it becomes crucial to assess the extent to which SUNet can effectively handle the presence of coloured Gaussian noise. Thus, our tests examine SUNet's generalisability in the presence of CGN.

\subsection{Debiasing with multi-resolution support}
\label{MRS}

\begin{figure}[h!]
    \centering
    \subfigure[\makebox{} ]
    {\label{subfig:MRSupp_2}\includegraphics[width=\linewidth]{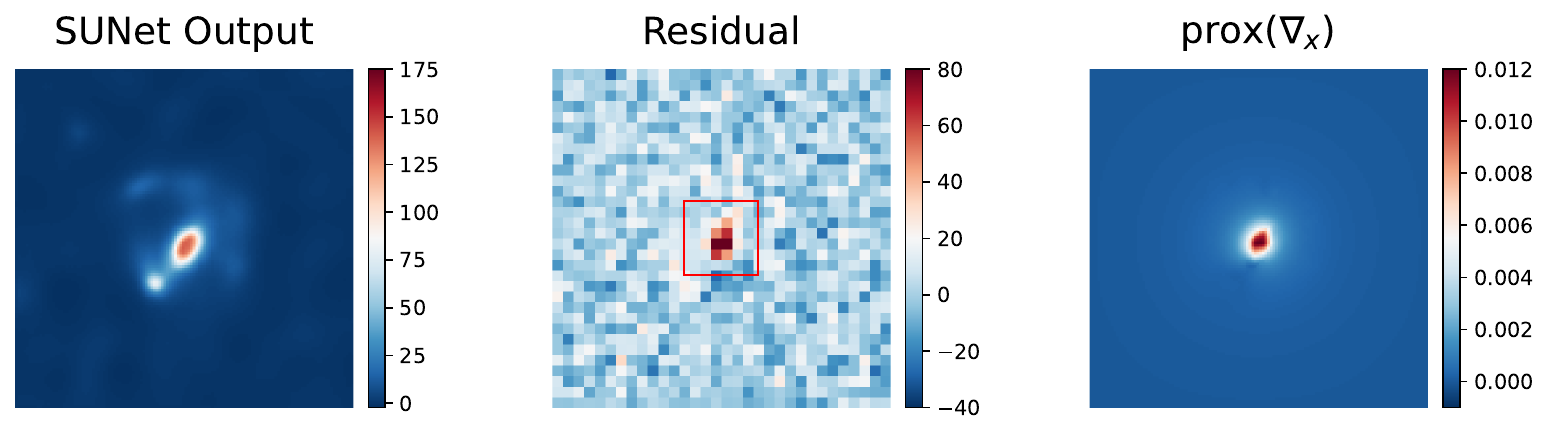}}\\
    
    \subfigure[\makebox{}]
    {\label{subfig:MRSupp_1}\includegraphics[width=0.95\linewidth]{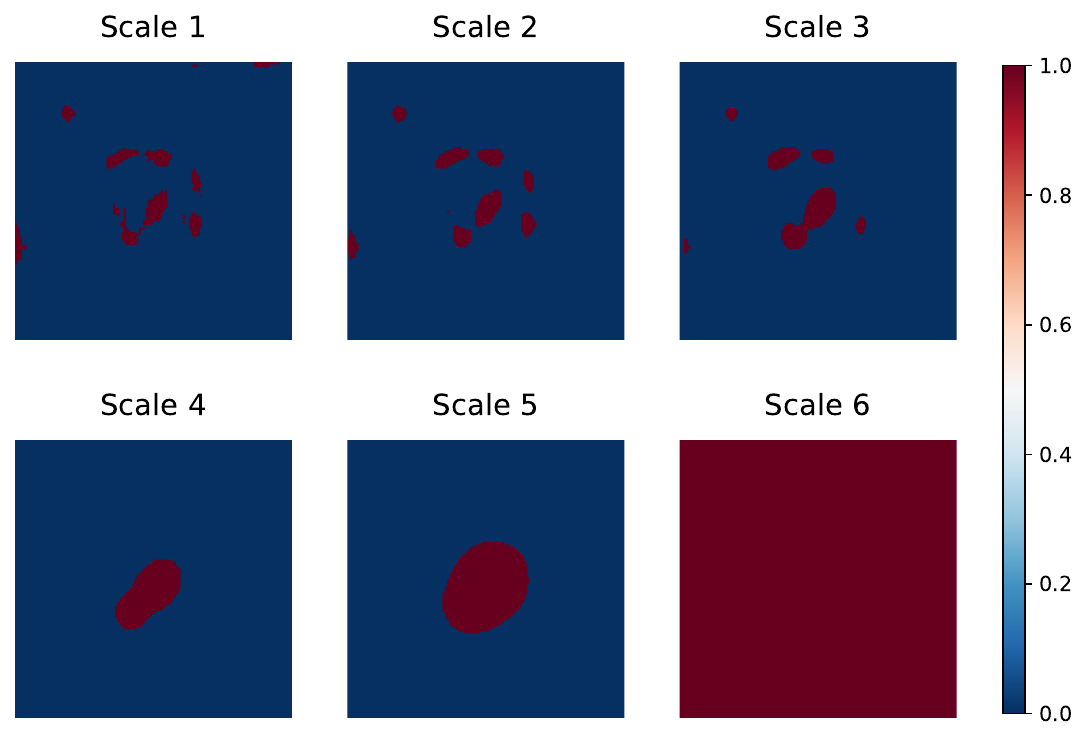}}\\

    \subfigure[\makebox{}]
    {\label{subfig:MRSupp_3}\includegraphics[width=\linewidth]{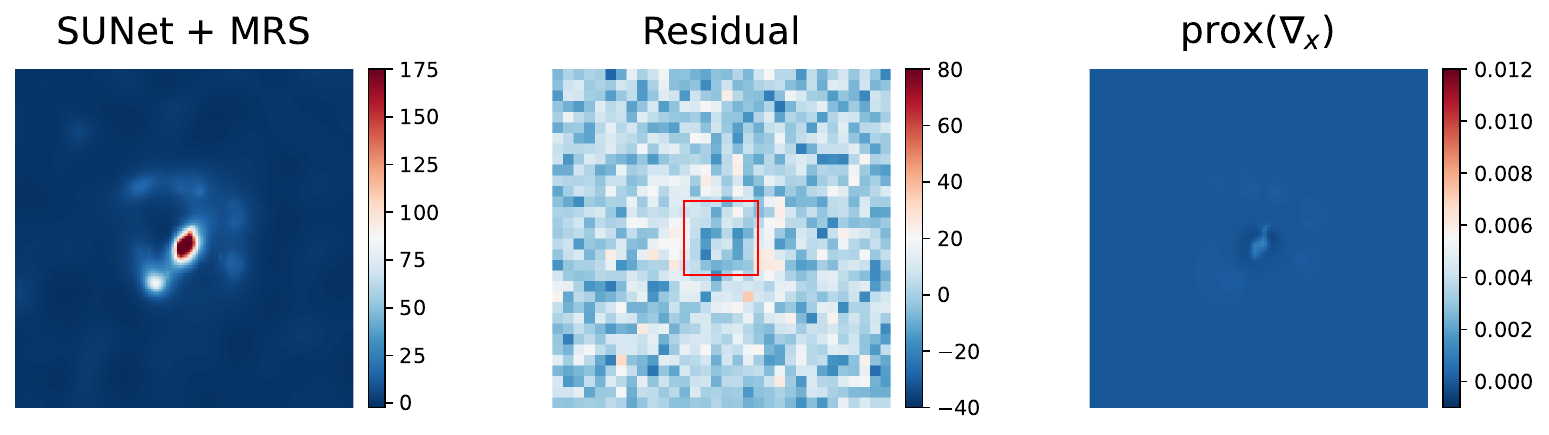}}\\   
    
    \subfigure[\makebox{}]
    {\label{subfig:MRSupp_4}\includegraphics[width=0.75\linewidth]{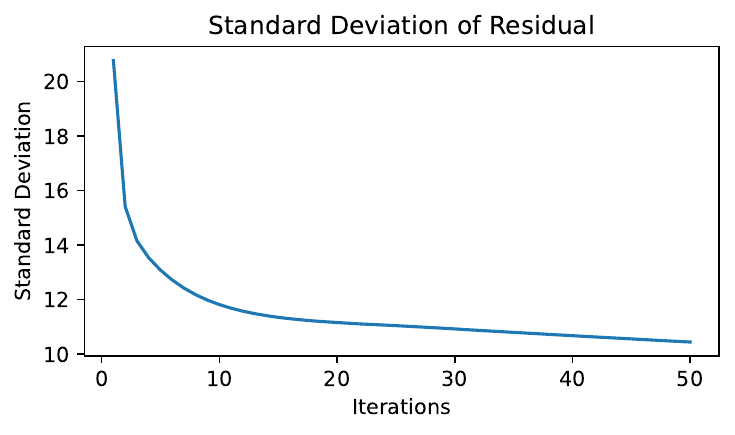}}\\    

    \caption{Iterative recovery of lost flux through debiasing using multi-resolution support. (\ref{subfig:MRSupp_2}): Original SUNet output. The red square highlights the residual flux lost. (\ref{subfig:MRSupp_1}): Multi-resolution support matrices at each decomposed scale. (\ref{subfig:MRSupp_3}): Debiased solution after iterative correction with multi-resolution support highlighting the reduction in structured residuals. (\ref{subfig:MRSupp_4}): Standard deviation of the residual within the highlighted region as a function of the number of iterations. The process was stopped upon achieving convergence.
    }
    \label{fig:MRSupp}
\end{figure}

Our study reveals that the deep learning-based solution introduces a bias, evident in the form of positive structures in the residuals, as illustrated in Figure \ref{subfig:MRSupp_2}. This bias can have implications on the accuracy of the scientific analyses, potentially influencing the flux estimation within the recovered features in the reconstructed images. In recovering the lost flux and enhancing image sharpness, multi-resolution support has been proven effective \citep{MRSupp}. Denoting the Starlet transform as $\Phi$ and the SUNet output solution as $\mathbf{x_0}$, an iterative process allowed for the recovery of flux from the residual $\mathbf{r_0}$. In each iteration, denoted as $\mathbf{j}$, a debiasing correction term was applied by multiplying the multi-resolution support matrices $\mathbf{M}$ of the SUNet output solution $\mathbf{x_0}$ at each scale with the Starlet decomposition of the gradient of the residual, incrementally modifying the deconvolved image:
\begin{equation}
    \mathbf{x_{j+1}} = \mathbf{x_j} + \text{prox}(\mathbf{\nabla_x}),
\end{equation}


\noindent where 
\vspace{1em}

\(\mathbf{r_j} =  \mathbf{y} - \mathbf{H x_j}\) \vspace{0.3em},

\(\mathbf{\nabla_x} = \mathbf{H^\top r_j}\) \vspace{0.3em},

\(\text{prox}(\mathbf{\nabla_x}) = (\Phi^\top \mathbf{M} \Phi) \mathbf{\nabla_x}\) \vspace{0.3em}, and

\(\mathbf{M} = \text{MRS}(\Phi(\mathbf{x_0}))\) .\\

The acronym ‘MRS’ stands for multi-resolution support and indicates a Boolean measure of whether an image $I_0$ contains information at a specific pixel and scale. If \(c\) represents the wavelet coefficient at a given scale and \(\lambda\) is the threshold value, the multi-resolution support operation can be defined as:
\begin{equation}
    \text{MRS}(c) =
    \begin{cases}
        1, & \text{if } |c| > \lambda \\
        0, & \text{otherwise}
    \end{cases}
\end{equation}

The flux-recovery process is illustrated in Figure \ref{fig:MRSupp}. The iterative process was stopped once convergence was achieved in the standard deviation of the residual as a function of the number of iterations, as seen in sub-figure \ref{subfig:MRSupp_4}. A more detailed study on the impact of debiasing with multi-resolution support on neural networks is presented in Appendix \ref{sec:MRS_appendix}.

\section{Dataset and experiments}
\label{sec:data_exp}

\subsection{Training dataset generation}
\label{datagen}

We extracted HST cutouts measuring $128 \times 128$ pixels from CANDELS \citep{CANDELS, CANDELS_HST} in the $F606W$ filter ($V$-band). These cutouts were then convolved with a Gaussian PSF having a full width at half maximum (FWHM) of 15 pixels. Following the convolution, we injected white Gaussian noise with a standard deviation denoted as $\mathbf{\sigma_{noise}}$. This choice ensured that the faintest object in our dataset had a peak signal-to-noise (S/N) close to one and was barely visible. With this particular value of $\mathbf{\sigma_{noise}}$, our dataset exhibited a range of S/N values depending on the magnitude of each galaxy. To standardise the images, each image was normalised within the $[-1,1]$ range. This normalization involved subtracting the image's mean and scaling the peak value by the image's maximum value. Finally, the batch of images was randomly divided into training-validation-test subsets in the ratio $0.8:0.1:0.1$.

\subsection{Training the SUNet}
\label{training}

The SUNet architecture was trained using a Titan RTX Turing GPU with 24 GB RAM for each job. The training aimed at learning the mapping from the Tikhonov output $\hat{\mathbf{x}}$ (containing CGN) to the corresponding HST image $\mathbf{x_t}$. The training utilised the Adam optimiser \citep{kingma2014adam} with an initial learning rate of $10^{-3}$, gradually halving every 25 epochs until reaching a minimum of $10^{-5}$. As in the original SUNet paper \citep{sunet}, $\ell_1$-loss was used for training. A more detailed discussion on the significance of the training loss function is given in Appendix \ref{sec:hallucintaions}. The input images were processed in mini-batches of size $16$. The dataset was augmented with random rotations in multiples of 90°; translations and flips were along horizontal and vertical axes. Starting with $22,317$ images in our initial training dataset, we increased its diversity by a factor of ten by performing augmentation.

\subsection{Test dataset}
\label{EDisCS}

We utilised the ESO Distant Cluster Survey \citep[EDisCS;][]{white2005} as our benchmark dataset to assess the performance of our deconvolution method. EDisCS is an extensive ESO Large Programme focused on the analysis of $20$ galaxy clusters within the redshift range $0.4 < z < 1$ and covering a diverse range of masses, with velocity dispersions ranging from approximately $200$ km s$^{-1}$ to $1000$ km s$^{-1}$. All of the clusters benefited from the deep $B$, $V$, $R$, and $I$ photometry obtained with FORS2 at the VLT. Additionally, a subset of ten clusters was imaged with the ACS at the HST in the F814W filter \citep{Simard_2002}. Previous work by \cite{cantale} employed the deconvolution technique Firedec to analyse spiral disc colours for EDisCS cluster galaxies, investigating trends with cluster masses and lookback time. For our study, we focused on analysing a subset of EDisCS clusters at three distinct redshifts—$z\approx0.58$, $z\approx0.7$, and $z\approx0.79$—using our proposed deconvolution method, and we went a step further by investigating the star-forming
clumps in these galaxies. Table \ref{tab:clusters} provides a summary of the properties of these clusters and the number of galaxies in each of them. 

\begin{table}[htbp]
    \centering
    \renewcommand{\arraystretch}{1.3}
        \caption{Summary of the EDisCS clusters considered for analysis.}
    \begin{tabular}{ccc}
    
        \hline
        \textbf{Cluster ID} & \textbf{$z_{cl}$} & \textbf{$N_{cl}$} \\
        \hline
        
        cl1037.9-1243 & 0.5805 & 11 \\
        \hline
        cl1040.7-1155 & 0.7043 &  19\\
        cl1054.4-1146 & 0.6972 & 20 \\
        cl1103.7-1245b & 0.7031 &  6\\
        \hline
        cl1216.8-1201 & 0.7955 & 28 \\
        \hline
    \end{tabular}
    \tablefoot{The clusters are grouped into three redshift categories: $z\approx0.58$, $z\approx0.70$, $z\approx0.79$. The term $z_{cl}$ indicates the cluster redshift, and $N_{cl}$ is the number of spectroscopically confirmed cluster galaxies.  }
    \label{tab:clusters}
\end{table}

Our analysis specifically considered the $V$ $(555 nm)$, $R$ $(655 nm)$, and $I$ $(768 nm)$ photometric bands, allowing us to evaluate the effectiveness of our deconvolution method in capturing variations across these photometric bands. To enable a more detailed analysis, we grouped these EDisCS galaxies based on their disc colour in order to study the different trends. Since the EDisCS clusters were solely observed in the F814W filter for HST, our deconvolution method presents a unique opportunity to extract high spatial resolution galaxy properties from the ground-based FORS2 multi-band observations. Furthermore, we demonstrate the superiority of SUNet over Firedec in generating cleaner deconvolved images, as detailed in Sections \ref{sec:dec_comparison} and \ref{sec:dec_vlt}.

\section{Results}
\label{results}

\begin{figure}[h!]
\begin{center}
\includegraphics[width=0.85\linewidth]{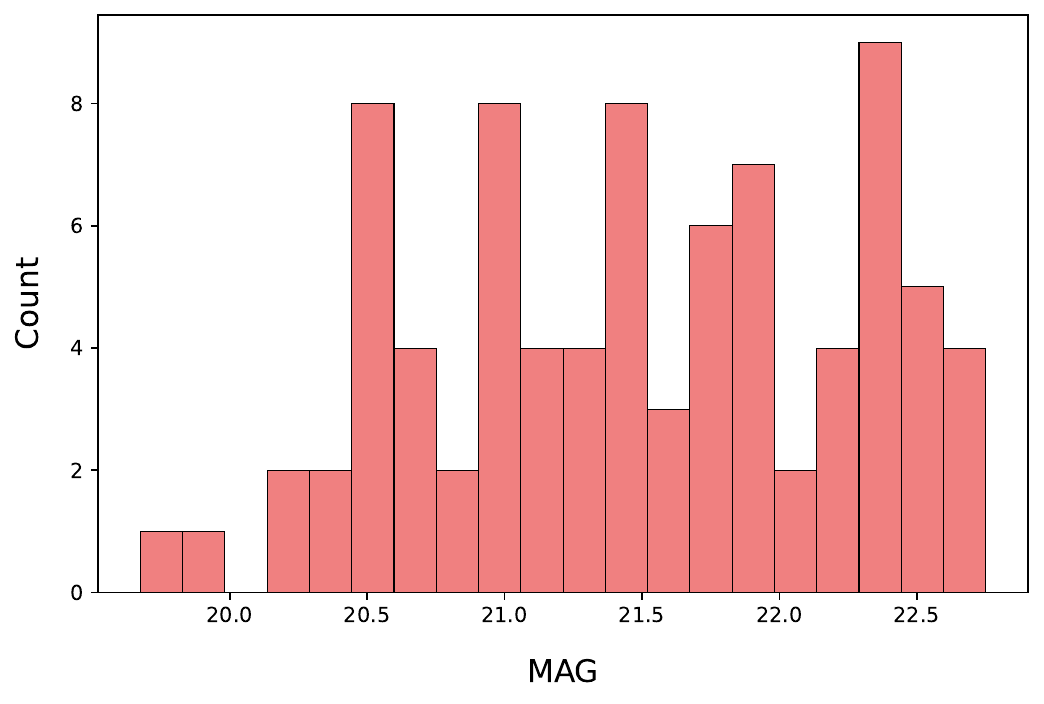}
\end{center}
\caption{\label{fig:hist_mag} {Distribution of the galaxy magnitudes in the HST F814W filter for the EDisCS samples, which were solely observed in the F814W filter for HST.}}
\end{figure}

The EDisCS images in the $V$-, $R$-, and $I$-bands served as inputs for the SUNet deconvolution framework, yielding corresponding deconvolved outputs. The algorithm operates with exceptional speed, requiring only $\approx 15.2$ ms to deconvolve a single image on a Titan RTX Turing GPU with 24 GB RAM—approximately $10^4$ times faster than Firedec, on average. A histogram of the magnitudes of the same objects in the HST F814W filter is shown in Figure \ref{fig:hist_mag}. In Section \ref{sec:size_clump}, we detail our approach to detecting sizes and the number of clumps in the deconvolved outputs. Section \ref{sec:dec_comparison} then presents a comparative analysis of the quality of our SUNet outputs against Firedec \citep{firedec}. With the validity of our method established, Section \ref{sec:dec_vlt} presents the deconvolved results and is followed by a thorough analysis in Section \ref{sec:EDisCS_analysis}.

\subsection{Objects size and clump detection}
\label{sec:size_clump}

\begin{figure}[h!]
    \centering
    
    \subfigure[\makebox{}]
    {\label{subfig:clump2}\includegraphics[width=\linewidth]{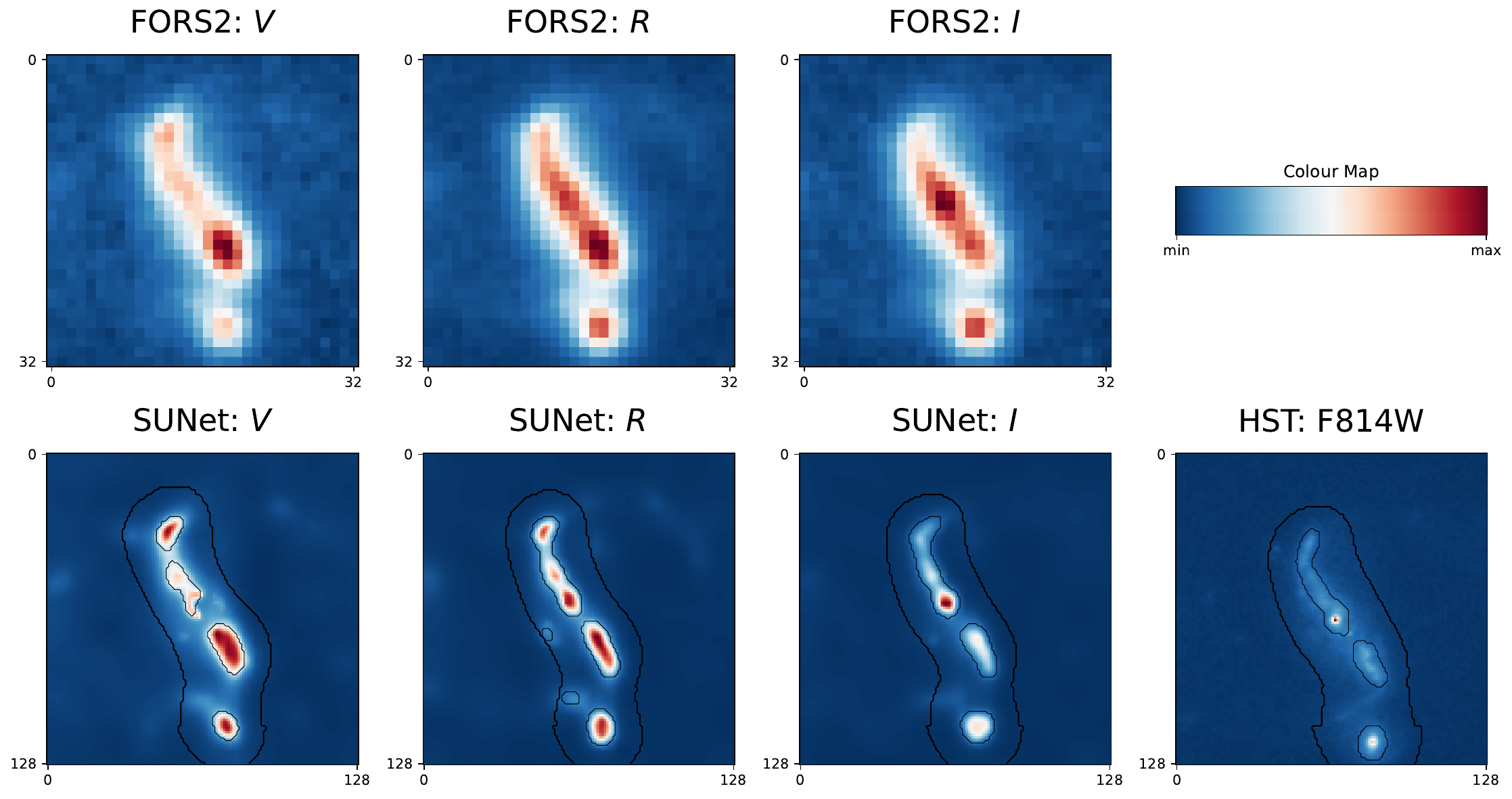}}\\

    \subfigure[\makebox{}]
    {\label{subfig:clump3}\includegraphics[width=\linewidth]{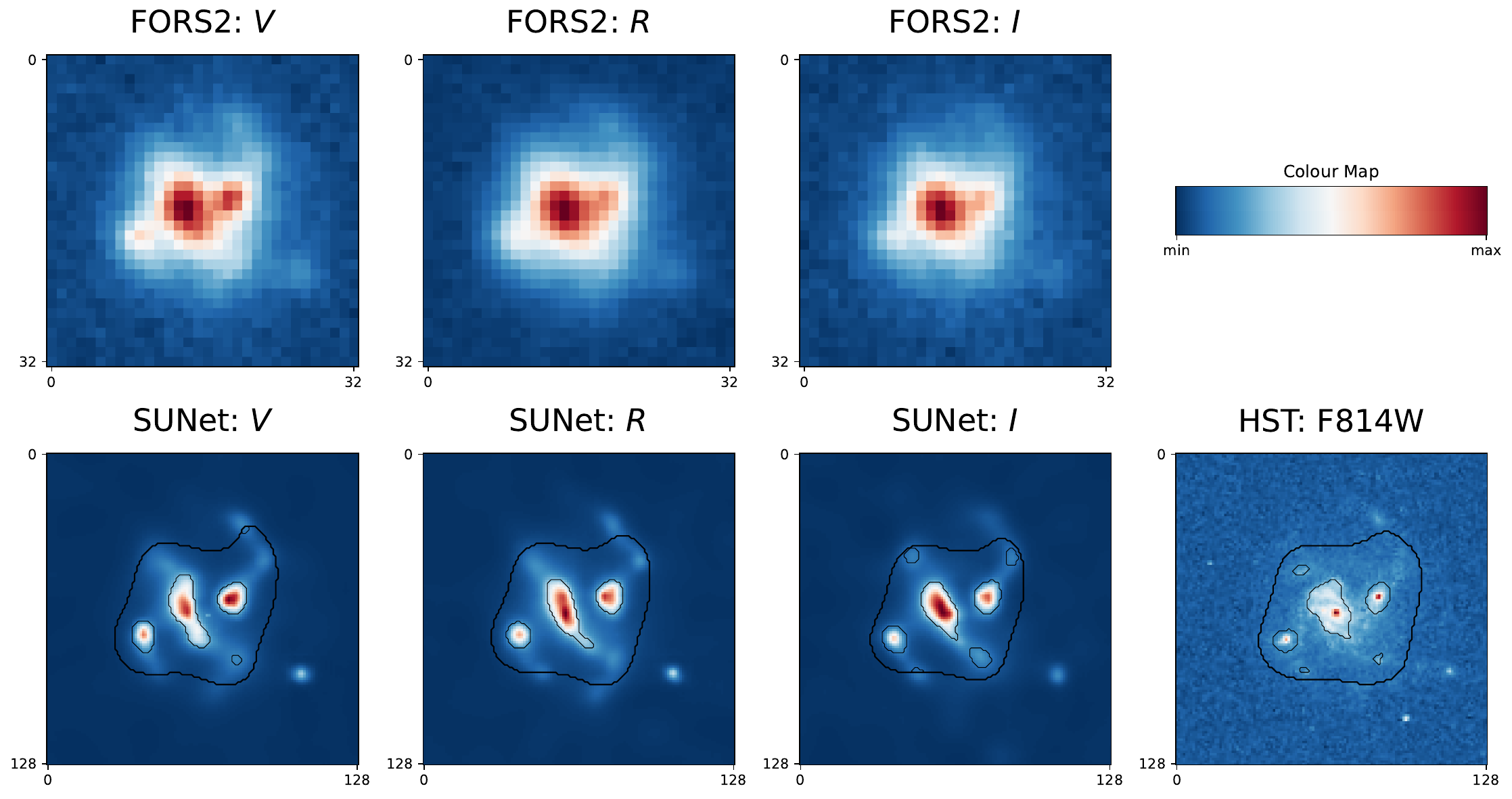}}\\
    
    \caption{Size detection (outer contour) and clump detection (inner contours) using SCARLET. The first row shows the FORS2 images in the $V$-, $R$-, and $I$-bands, with the corresponding SUNet outputs displayed directly below. For comparison, the HST image in the F814W filter is shown adjacent to the SUNet $I$-band output. All images are decomposed into five scales, with the fourth scale chosen for size detection and the second scale for clump detection}
    \label{fig:size_clump}
\end{figure}

To detect clumps, or small-scale structures, and measure the sizes of galaxies, we employed the SCARLET Python package \citep{MELCHIOR2018129}. Utilising the Starlet transform from SCARLET, our approach involved decomposing images into different scales, where each scale captures a specific frequency component. To ensure consistency for an unbiased comparison, we maintained fixed algorithm parameters across different bands and objects. All images underwent decomposition into five scales, with the fourth scale chosen for size detection and the second scale for clump detection. A $5\sigma$ detection threshold was applied to each scale during the process. In the context of clump detection, the Starlet transform was computed using the standard deviation solely within the region enclosed by the size detection outline, rather than considering the entire image. This refined approach ensured more precise thresholding in the Starlet space. Finally, a clump was only considered valid if it lay within the size detection outline, ensuring that background artefacts were excluded. In Figure \ref{fig:size_clump}, we present examples of size detection and their corresponding clump detection cases.

\subsection{Comparison with classical methods}
\label{sec:dec_comparison}

\begin{figure}[h!]
\begin{center}
\includegraphics[width=0.95\columnwidth]{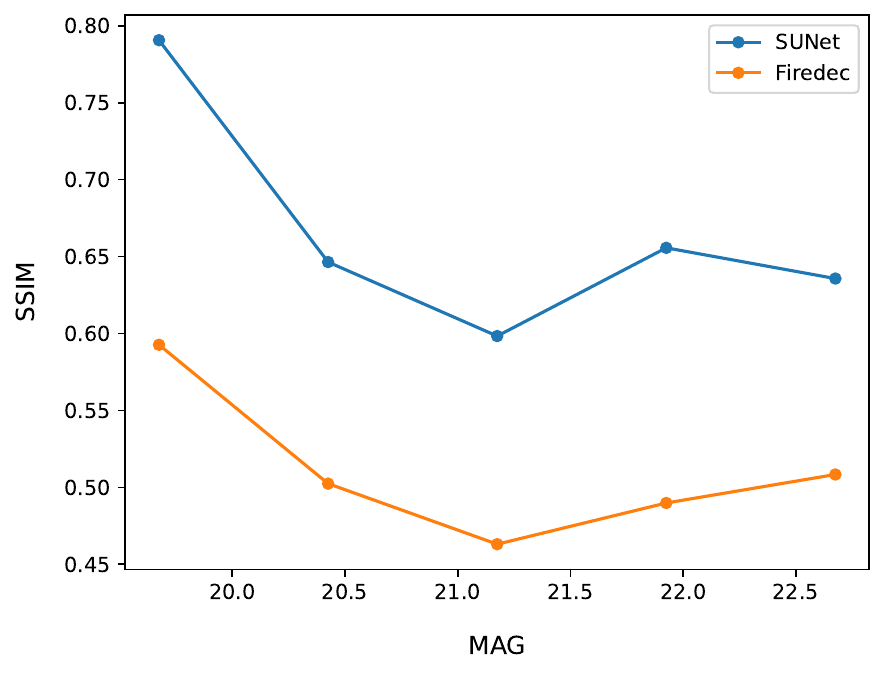}
\end{center}
\caption{\label{fig:ssim} {SSIM between the $I$-band deconvolved outputs and the HST images in the F814W filter as a function of object magnitude. An SSIM of one implies identical images.}}
\end{figure}

\begin{figure}[h!]
\begin{center}
\includegraphics[width=0.95\columnwidth]{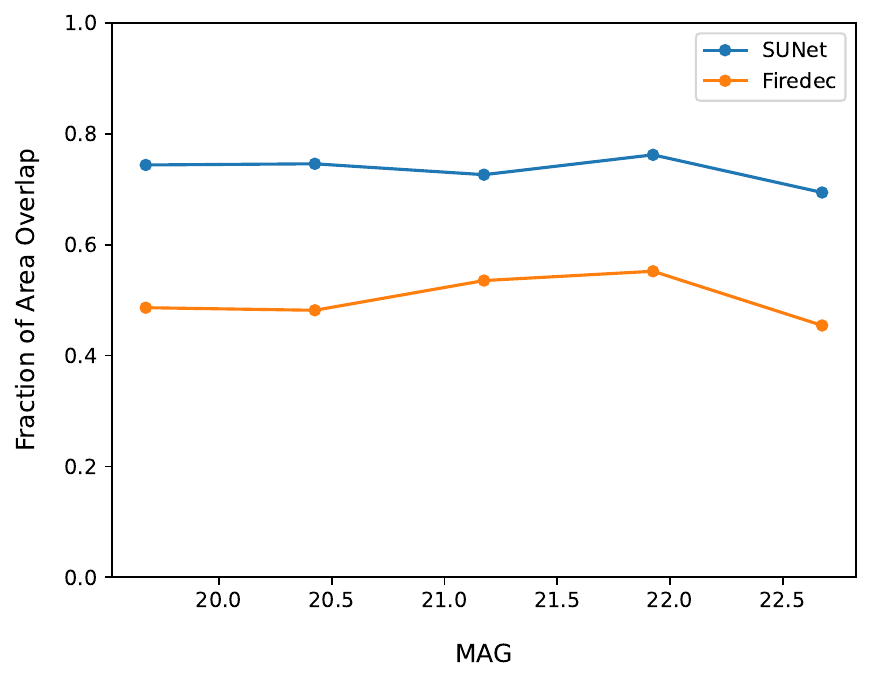}
\end{center}
\caption{\label{fig:overlap} {Fraction of area overlap between the small-scale structure detections of the $I$-band deconvolved outputs and HST images in the F814W filter.}}
\end{figure}

We conducted a thorough performance comparison between SUNet and Firedec, a classical deconvolution method based on wavelet regularisation \citep{firedec}. For direct comparison with HST quality, we concentrated on the $I$-band outputs for each method, as the EDisCS clusters were exclusively observed in the F814W filter for HST. Both methods exhibited a better performance on low-magnitude (or high-S/N) images, with a gradual decline in performance in the high-magnitude (or low-S/N) regime. In this case, the mean squared error metric between the deconvolved outputs and the ground truth HST images is not a robust metric for indicating similarity since it is biased by the background noise in the HST images \citep{mse}. Instead, we used the structural similarity index measure (SSIM), a full reference metric which quantifies the similarity between two images by comparing their structural information or spatial interdependencies \citep{ssim}. An SSIM of one implies identical images. The observed trends are illustrated in Figure \ref{fig:ssim}. We further assessed the ability of the deconvolution algorithms to accurately resolve small-scale structures. For this, we leveraged the SCARLET Python package, as detailed in Section \ref{sec:size_clump}. The fraction of area overlap between small-scale structure detections in the deconvolved outputs and HST is depicted in Figure \ref{fig:overlap}. Based on these metrics, SUNet clearly outperforms Firedec.

\begin{figure}[h!]
    \centering
    
    \subfigure[\makebox{}]
    {\label{subfig:comp2}\includegraphics[width=\linewidth]{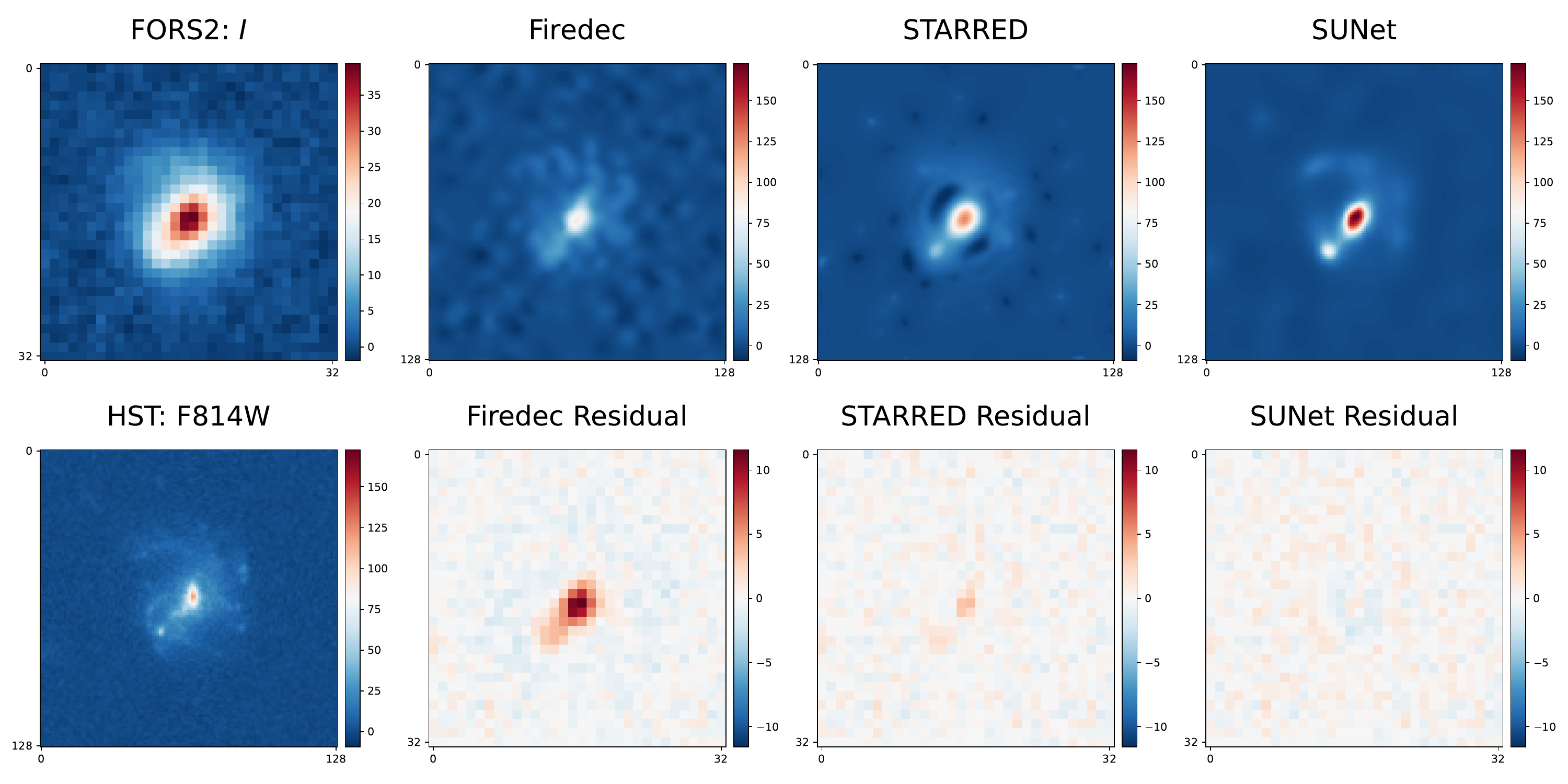}}\\

    \subfigure[\makebox{}]
    {\label{subfig:comp3}\includegraphics[width=\linewidth]{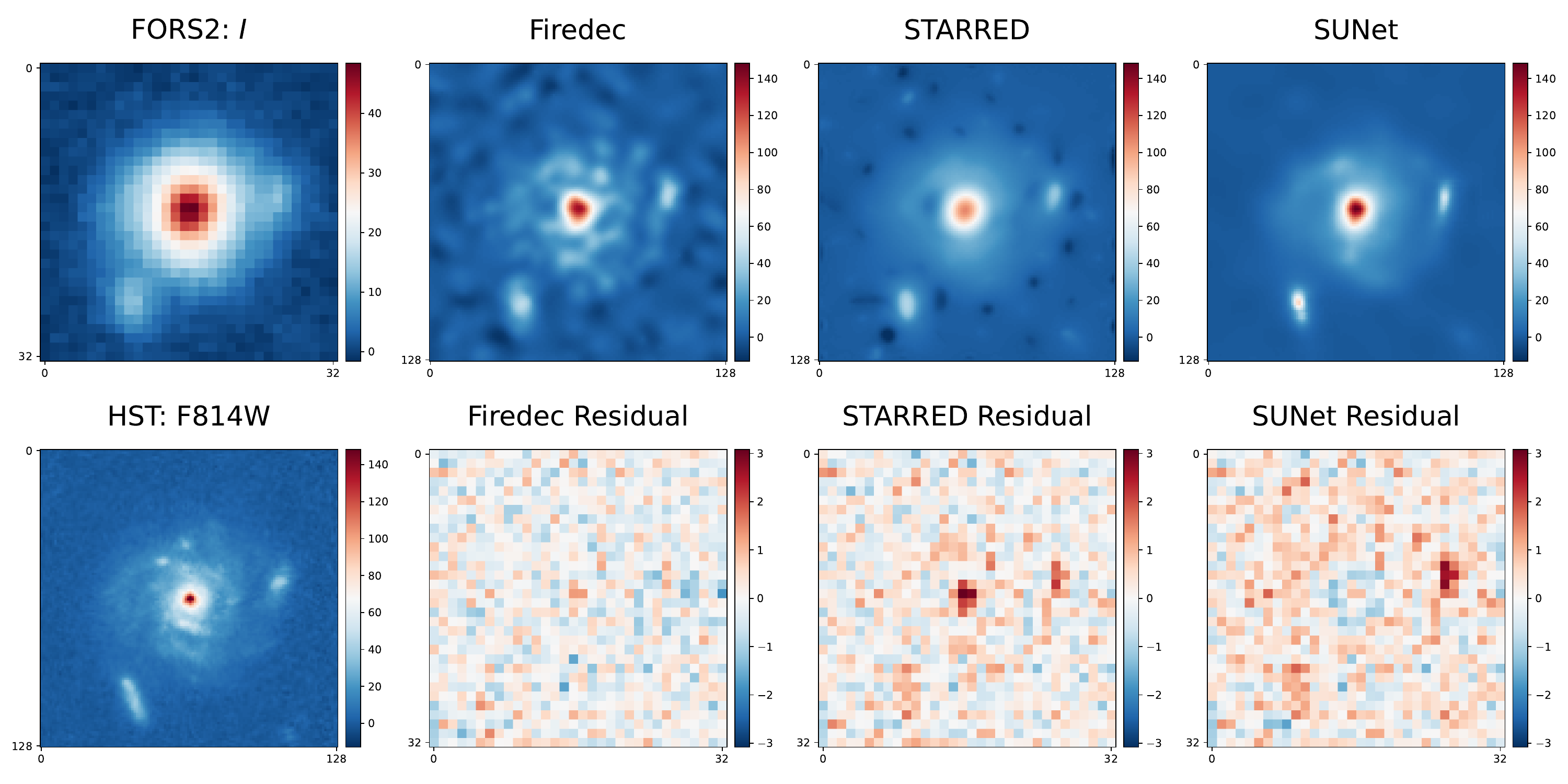}}\\  

    \subfigure[\makebox{}]
    {\label{subfig:comp4}\includegraphics[width=\linewidth]{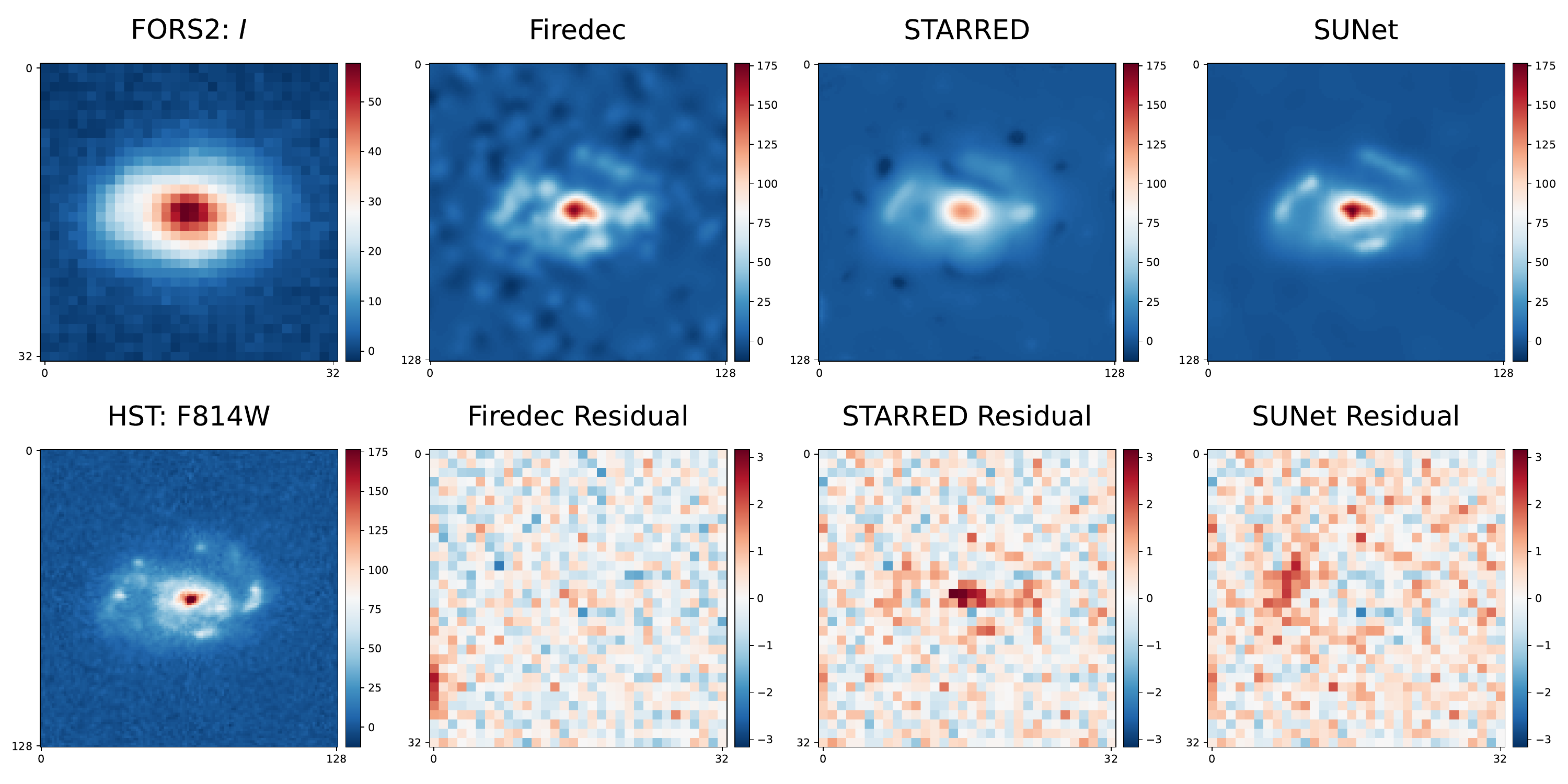}}\\  

    
    \caption{Visual comparison between the deconvolved outputs. The FORS2 image in the $I$-band is displayed in the top-left corner, with the HST image in the F814W filter directly below it. The Firedec, STARRED, and SUNet images in the $I$-band are shown in the second, third, and fourth columns of the first row, respectively. Beneath each output, the corresponding residual is depicted, which is defined as follows: residual $=$ noisy VLT image $-$ PSF $\ast$ deconvolved image.}
    \label{fig:comp}
\end{figure}

Building on the foundation of Firedec, an enhanced method named STARRED was recently introduced by \cite{STARRED}. STARRED brings innovation by incorporating an isotropic wavelet basis known as Starlets \citep{Starck2015} that can refine the regularisation process when solving the deconvolution problem. The outputs of Firedec, STARRED, and SUNet are shown in Figure \ref{fig:comp}. Upon visual inspection of the outputs and the residuals (residual $=$ noisy image $-$ PSF $\ast$ deconvolved image), it was evident that SUNet consistently generalises better than Firedec and STARRED.

\subsection{SUNet deconvolution results}
\label{sec:dec_vlt}

\begin{figure}[h!]
    \centering
    \subfigure[\makebox{}]
    {\label{subfig:sunet1}\includegraphics[width=\linewidth]{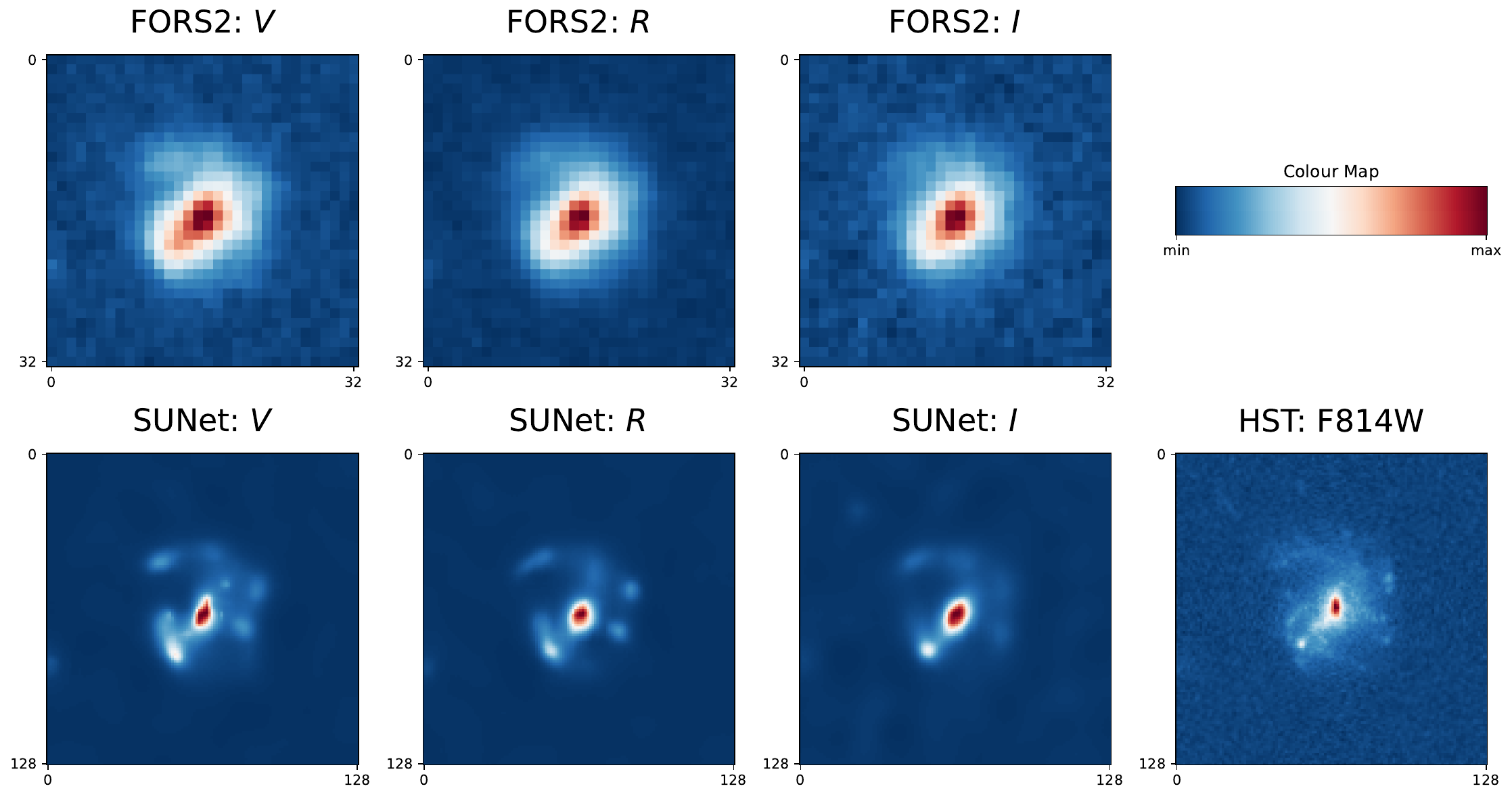}}\\
    
    \subfigure[\makebox{}]
    {\label{subfig:sunet2}\includegraphics[width=\linewidth]{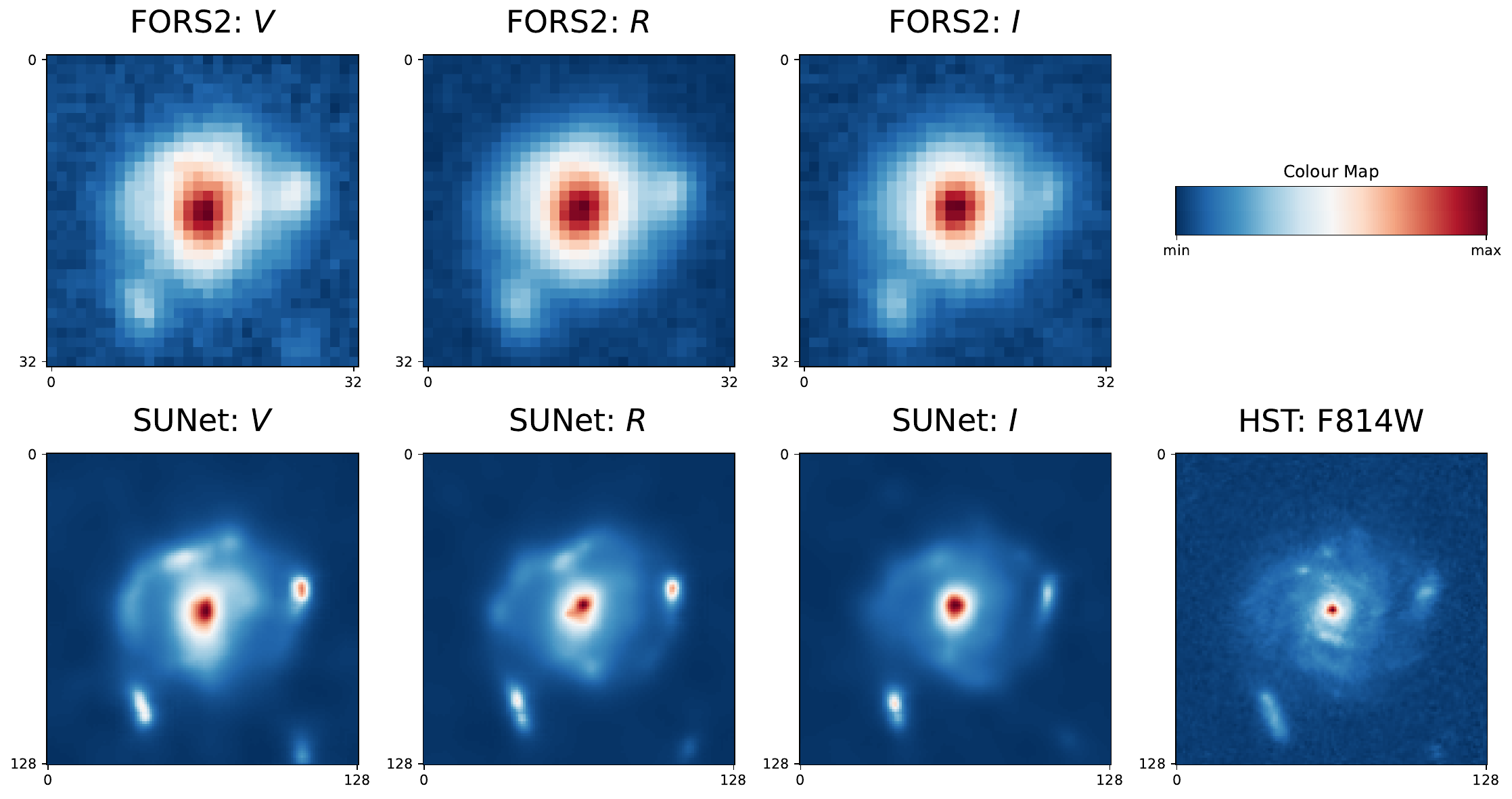}}\\

    \subfigure[\makebox{}]
    {\label{subfig:sunet3}\includegraphics[width=\linewidth]{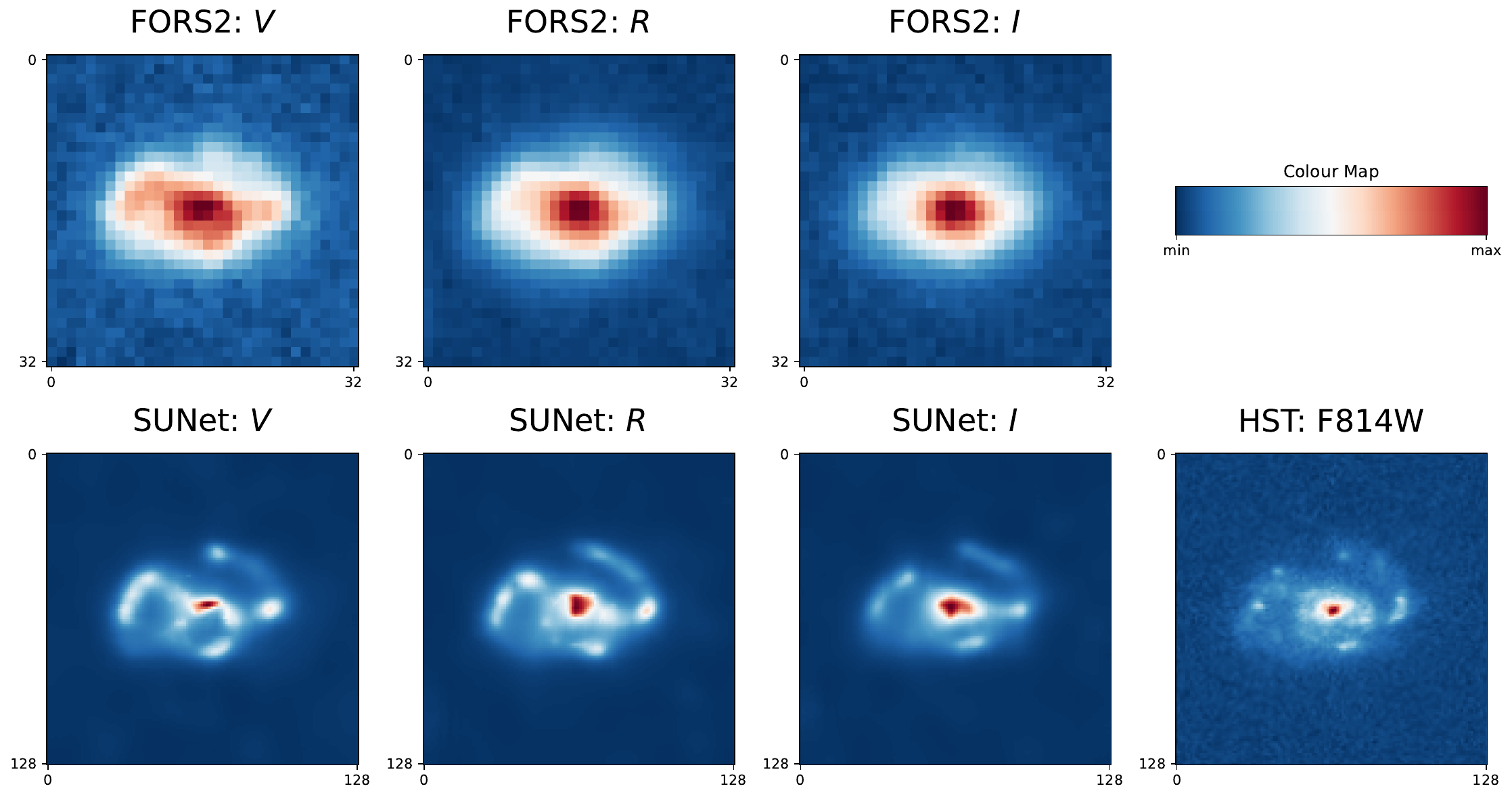}}\\  
    
    \caption{Images of a few SUNet outputs without clump and size detection outlines, emphasising the accuracy in recovering the shapes of galaxies. The first row shows the FORS2 images in the $V$-, $R$-, and $I$-bands, with the corresponding SUNet outputs displayed directly below. For comparison, the HST image in the F814W filter is shown adjacent to the SUNet $I$-band output.}
    \label{fig:sunet}
\end{figure}

\begin{figure}[h!]
\begin{center}
\includegraphics[width=\columnwidth]{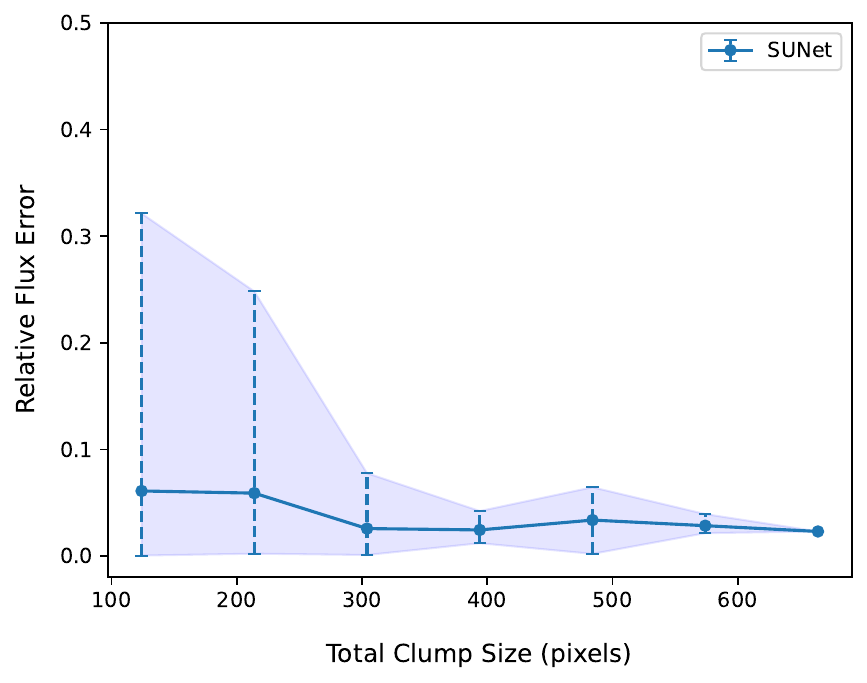}
\end{center}
\caption{\label{fig:flux_err} {Relative flux error between the SUNet $I$-band deconvolved outputs and HST images in the F814W filter. Each data point in the plot represents the mean value for a specific bin, while the error bars depict the upper and lower bounds within which $95\%$ of the data points fall.}}
\end{figure}

The methods were put to the test using real ground-based images captured by the FORS2 camera at VLT in Chile. Notably, SUNet showed a remarkable ability to effectively generalise to images with entirely different noise properties than those present in the training dataset, as depicted in Figure \ref{fig:sunet}. As illustrated, we were able to successfully recover the morphology and lost small-scale structures. Figure \ref{fig:flux_err} presents the trend in the relative flux error between the SUNet deconvolved outputs and the corresponding HST targets as a function of the total clump size. The uncertainty in the flux level is higher for smaller clumps. 

\subsubsection{Resolution recovery}
\label{res_recovery}

To gauge the achieved resolution in the deconvolved outputs, we calculated the average ratio between the areas of the smallest detected clump in the SUNet output and its counterpart in the HST image. This ratio, approximately $2.58$, implies an average SUNet output resolution of around $0.129 \arcsec$, considering the known HST resolution is $0.05 \arcsec$.

\subsubsection{False positives and false negatives}
\label{fp+fn}

To assess the reliability of our deconvolution method for real-world applications, we conducted an analysis to estimate the number of false positives and false negatives in our study. We ran SCARLET to detect clumps in both SUNet outputs and HST ground truths. For each HST clump, we checked whether the SUNet-identified clump centroid fell within a $5$-pixel radius of the HST clump centroid. Clumps failing this criterion were considered false positives. The false positive rate was computed across our entire EDisCS dataset by tallying the total count of false positives and dividing it by the overall count of detected clumps in the HST images. However, it is important to note that this result may be biased due to SCARLET's performance. Instances exist where SCARLET identifies a clump in the SUNet image but misses it in the corresponding HST image, leading to an elevated false positive count. To address this, we employed visual inspection to filter out falsely detected cases. The resultant false positive rate was determined to be approximately $4.16\%$. Using a similar approach, we also computed the false negative rate, indicating the probability of missing clumps in the deconvolved images that are present in the HST image. This rate was found to be 
$3.57\%$, signifying a very low probability of missing features. To obtain a more statistically robust evaluation, we tested the method on another dataset of $2232$ galaxies extracted from CANDELS \citep{CANDELS} and compared it with other neural networks. This work is shown in Appendix \ref{sec:hallucintaions}.

\subsection{Analysis of deconvolved EDisCS galaxies}
\label{sec:EDisCS_analysis}

As a sanity check, we verified that any conclusions drawn regarding the internal properties of our sample galaxies are not influenced by biases in their sizes, either in relation to redshift or disc colours.  To this end, a histogram of galaxy sizes grouped by their parent cluster redshift and disc colour is presented in Appendix \ref{sec:supp_fig}. Both plots show that all galaxies have the same global spatial extent.

\begin{figure*}[t!]
\begin{center}
\includegraphics[width=\textwidth]{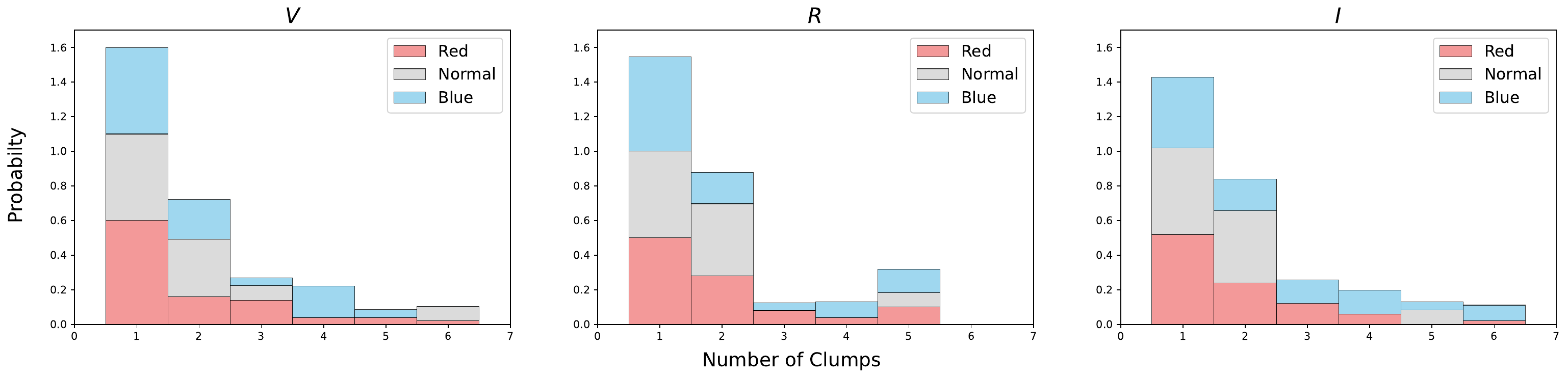}
\end{center}
\caption{\label{fig:hist_clumps_color} {Histogram of the number of clumps in galaxies in the $V$-, $R$-, and $I$-bands grouped by their parent disc colour. Each coloured bar in the plots corresponds to a specific disc colour, and the bars for different disc colours are stacked on top of each other. Galaxies are classified as `Red' if they are redder, `Normal' if they are comparable, and `Blue' if they are bluer than the field members.}}
\end{figure*}

Following \cite{cantale}, our analysis focuses on the population of galaxies with discs redder than their field counterparts. Some of our sample galaxies have normal colours and hence fall into the colour distribution of the field galaxies, and some are bluer (by more than $1\sigma$ of the colour distribution). A few known physical processes can induce enhanced star-forming activity, but we were instead interested in the possible evidence for quenching mechanisms. Therefore, in the latter case, normal and blue disc galaxies form a common broad class of systems to which we compared the redder ones. Employing the clump detection method outlined in Section \ref{sec:size_clump}, we computed the histogram of the number of clumps in galaxies, categorised by disc colour, as shown in Figure \ref{fig:hist_clumps_color}. We note that the only-one-clump case reflects the identification of the central and luminous part (bulge) of the galaxies. As illustrated in Figures \ref{fig:size_clump} and \ref{fig:sunet}, clumps in the $V$-band are brighter than those in the $R$- and $I$-bands, in agreement with the spectral energy distribution of young stellar populations.  In principle, it is therefore easier to detect clumps in the $V$-band at equivalent photometric depths of the images. This may explain the more continuous distribution in the number of clumps, from one to six, in the $V$-band. Even so, as we witnessed in Figure \ref{fig:hist_clumps_color}, the general trend is the same from one band to the other. This trend is clear and reveals that the red discs that were initially identified by \citet{cantale} have fewer clumps than their bluer counterparts, most likely due to an earlier cessation of star formation. This result opens promising prospects for future studies on larger samples and over larger look intervals.



\section{Conclusion}
\label{concl}
We have proposed a deconvolution framework involving a two-step process—namely, Tikhonov deconvolution and post-processing with an SUNet denoiser—and an additional debiasing step using multi-resolution support. SUNet was trained on galaxy images from the CANDELS survey and demonstrated superior performance compared to Firedec in the astrophysical context. After establishing the validity of our method, we applied it to deconvolve a set of galaxies from the EDisCS cluster at three different redshifts. Using SCARLET, we provided further analysis of the galaxies in terms of their size and disc colour. We quantified the number of clumps in these galaxies, examining their relationship with disc colour. 
Our results, based on both quantitative metrics and visual assessments, highlight the effectiveness of SUNet and showcase its ability to generalise unseen real images with diverse noise properties, which can be attributed to its transformer-based backbone involving the self-attention mechanism. 

In summary, this work introduces and evaluates an advanced deconvolution framework applied to ground-based astronomical images. The key findings and contributions include the following:

\begin{itemize}
    \item Resolution recovery: Based on our SCARLET detection procedure, SUNet demonstrates the capability to recover small-scale structures, with an average resolution of approximately $0.129 \arcsec$, and it outperforms classical algorithms such as STARRED and Firedec (Section \ref{res_recovery}).
    \item Generalisation to diverse noise properties: The method showcases robust generalisation to noise properties different from its training dataset, indicating its adaptability to various observational conditions.
    \item Clump analysis:
    Red discs exhibit fewer clumps than their bluer counterparts, affirming the lower presence of star-forming regions.
    \item False positive and false negative rates: Based on our SCARLET detection analysis on EDisCS, SUNet maintains a false positive rate of $4.16\%$ and a false negative rate of $3.57\%$, ensuring reliable feature recovery (Section \ref{fp+fn}). 
    \item Computational efficiency: The AI-based framework proves to be highly efficient, with an execution time of approximately $15.2$ ms per image, making it around $10^4$ times faster than traditional deconvolution methods such as STARRED and Firedec.

\end{itemize}

Our proposed technique can therefore be used with ground-based images to efficiently identify structures in the distant universe at high spatial resolution. The technique's applicability to multi-band observations further enhances its utility in studying various astrophysical phenomena. The efficiency of SUNet in processing large datasets and accelerating the deconvolution process opens up opportunities for swift analyses. Access to such a fast and robust deconvolution framework holds the potential to facilitate numerous astrophysical investigations.

\section{Data availability}
\label{sec:reproducible_research}

For the sake of reproducible research, the codes and the trained models used for this article are publicly available online. 

\begin{enumerate}
    \item The ready-to-use version of our deconvolution method\footnote{\url{https://github.com/utsav-akhaury/SUNet/tree/main/Deconvolution}}.
  \item The repository fork of the SUNet code used for training the network\footnote{\url{https://github.com/utsav-akhaury/SUNet}}.
  \item Link to the trained network weights\footnote{\url{https://doi.org/10.5281/zenodo.10287213}}.
  \item The repository forks of the Learnlet and Unet-64 codes used for comparison in the Appendices \ref{sec:MRS_appendix} and \ref{sec:hallucintaions}\footnote{\url{https://github.com/utsav-akhaury/understanding-unets/tree/candels}}.
\end{enumerate}

\begin{acknowledgements}
      This work was funded by the Swiss National Science Foundation (SNSF) under the Sinergia grant number CRSII5\_198674. This work was supported by the TITAN ERA Chair project (contract no. 101086741) within the Horizon Europe Framework Program of the European Commission, and the  Agence Nationale de la Recherche (ANR-22-CE31-0014-01 TOSCA). The authors thank David Donoho for useful discussions. 

\end{acknowledgements}

%
%

\bibliographystyle{aa} 
\bibliography{references.bib} 

\begin{thebibliography}{45}
\expandafter\ifx\csname natexlab\endcsname\relax\def\natexlab#1{#1}\fi

\bibitem[{Akhaury {et~al.}(2022)Akhaury, Starck, Jablonka, Courbin, \&
  Michalewicz}]{akhaury2022}
Akhaury, U., Starck, J.-L., Jablonka, P., Courbin, F., \& Michalewicz, K. 2022,
  Frontiers in Astronomy and Space Sciences, 9

\bibitem[{Cantale {et~al.}(2016{\natexlab{a}})Cantale, Courbin, Tewes,
  Jablonka, \& Meylan}]{firedec}
Cantale, N., Courbin, F., Tewes, M., Jablonka, P., \& Meylan, G.
  2016{\natexlab{a}}, A\&A, 589, A81

\bibitem[{Cantale {et~al.}(2016{\natexlab{b}})Cantale, Jablonka, Courbin,
  Rudnick, Zaritsky, Meylan, Desai, De~Lucia, Arag\'on-Salamanca, Poggianti,
  Finn, \& Simard}]{cantale}
Cantale, N., Jablonka, P., Courbin, F., {et~al.} 2016{\natexlab{b}}, A\&A, 589,
  A82

\bibitem[{Dosovitskiy {et~al.}(2021)Dosovitskiy, Beyer, Kolesnikov,
  Weissenborn, Zhai, Unterthiner, Dehghani, Minderer, Heigold, Gelly,
  Uszkoreit, \& Houlsby}]{vit}
Dosovitskiy, A., Beyer, L., Kolesnikov, A., {et~al.} 2021, in International
  Conference on Learning Representations

\bibitem[{{Euclid Collaboration} {et~al.}(2022){Euclid Collaboration},
  {Scaramella}, {Amiaux}, {Mellier}, {Burigana}, {Carvalho}, {Cuillandre}, {Da
  Silva}, {Derosa}, {Dinis}, {Maiorano}, {Maris}, {Tereno}, {Laureijs},
  {Boenke}, {Buenadicha}, {Dupac}, {Gaspar Venancio}, {G{\'o}mez-{\'A}lvarez},
  {Hoar}, {Lorenzo Alvarez}, {Racca}, {Saavedra-Criado}, {Schwartz}, {Vavrek},
  {Schirmer}, {Aussel}, {Azzollini}, {Cardone}, {Cropper}, {Ealet}, {Garilli},
  {Gillard}, {Granett}, {Guzzo}, {Hoekstra}, {Jahnke}, {Kitching}, {Maciaszek},
  {Meneghetti}, {Miller}, {Nakajima}, {Niemi}, {Pasian}, {Percival},
  {Pottinger}, {Sauvage}, {Scodeggio}, {Wachter}, {Zacchei}, {Aghanim},
  {Amara}, {Auphan}, {Auricchio}, {Awan}, {Balestra}, {Bender}, {Bodendorf},
  {Bonino}, {Branchini}, {Brau-Nogue}, {Brescia}, {Candini}, {Capobianco},
  {Carbone}, {Carlberg}, {Carretero}, {Casas}, {Castander}, {Castellano},
  {Cavuoti}, {Cimatti}, {Cledassou}, {Congedo}, {Conselice}, {Conversi},
  {Copin}, {Corcione}, {Costille}, {Courbin}, {Degaudenzi}, {Douspis},
  {Dubath}, {Duncan}, {Dusini}, {Farrens}, {Ferriol}, {Fosalba}, {Fourmanoit},
  {Frailis}, {Franceschi}, {Franzetti}, {Fumana}, {Gillis}, {Giocoli},
  {Grazian}, {Grupp}, {Haugan}, {Holmes}, {Hormuth}, {Hudelot}, {Kermiche},
  {Kiessling}, {Kilbinger}, {Kohley}, {Kubik}, {K{\"u}mmel}, {Kunz},
  {Kurki-Suonio}, {Lahav}, {Ligori}, {Lilje}, {Lloro}, {Mansutti}, {Marggraf},
  {Markovic}, {Marulli}, {Massey}, {Maurogordato}, {Melchior}, {Merlin},
  {Meylan}, {Mohr}, {Moresco}, {Morin}, {Moscardini}, {Munari}, {Nichol},
  {Padilla}, {Paltani}, {Peacock}, {Pedersen}, {Pettorino}, {Pires}, {Poncet},
  {Popa}, {Pozzetti}, {Raison}, {Rebolo}, {Rhodes}, {Rix}, {Roncarelli},
  {Rossetti}, {Saglia}, {Schneider}, {Schrabback}, {Secroun}, {Seidel},
  {Serrano}, {Sirignano}, {Sirri}, {Skottfelt}, {Stanco}, {Starck},
  {Tallada-Cresp{\'\i}}, {Tavagnacco}, {Taylor}, {Teplitz}, {Toledo-Moreo},
  {Torradeflot}, {Trifoglio}, {Valentijn}, {Valenziano}, {Verdoes Kleijn},
  {Wang}, {Welikala}, {Weller}, {Wetzstein}, {Zamorani}, {Zoubian}, {Andreon},
  {Baldi}, {Bardelli}, {Boucaud}, {Camera}, {Di Ferdinando}, {Fabbian},
  {Farinelli}, {Galeotta}, {Graci{\'a}-Carpio}, {Maino}, {Medinaceli}, {Mei},
  {Neissner}, {Polenta}, {Renzi}, {Romelli}, {Rosset}, {Sureau}, {Tenti},
  {Vassallo}, {Zucca}, {Baccigalupi}, {Balaguera-Antol{\'\i}nez}, {Battaglia},
  {Biviano}, {Borgani}, {Bozzo}, {Cabanac}, {Cappi}, {Casas}, {Castignani},
  {Colodro-Conde}, {Coupon}, {Courtois}, {Cuby}, {de la Torre}, {Desai},
  {Dole}, {Fabricius}, {Farina}, {Ferreira}, {Finelli}, {Flose-Reimberg},
  {Fotopoulou}, {Ganga}, {Gozaliasl}, {Hook}, {Keihanen}, {Kirkpatrick},
  {Liebing}, {Lindholm}, {Mainetti}, {Martinelli}, {Martinet}, {Maturi},
  {McCracken}, {Metcalf}, {Morgante}, {Nightingale}, {Nucita}, {Patrizii},
  {Potter}, {Riccio}, {S{\'a}nchez}, {Sapone}, {Schewtschenko}, {Schultheis},
  {Scottez}, {Teyssier}, {Tutusaus}, {Valiviita}, {Viel}, {Vriend}, \&
  {Whittaker}}]{Euclid2}
{Euclid Collaboration}, {Scaramella}, R., {Amiaux}, J., {et~al.} 2022, \aap,
  662, A112

\bibitem[{Fan {et~al.}(2022)Fan, Liu, \& Liu}]{sunet}
Fan, C.-M., Liu, T.-J., \& Liu, K.-H. 2022, in 2022 {IEEE} International
  Symposium on Circuits and Systems ({ISCAS}) ({IEEE})

\bibitem[{Grogin {et~al.}(2011)Grogin, Kocevski, \& Faber}]{CANDELS}
Grogin, N.~A., Kocevski, D.~D., \& Faber, S.~M. 2011, The Astrophysical Journal
  Supplement Series, 197, 35

\bibitem[{Guan {et~al.}(2020)Guan, Khan, Sikdar, \& Chitnis}]{fdunet}
Guan, S., Khan, A.~A., Sikdar, S., \& Chitnis, P.~V. 2020, IEEE Journal of
  Biomedical and Health Informatics, 24, 568

\bibitem[{{Guo} {et~al.}(2015){Guo}, {Ferguson}, {Bell}, {Koo}, {Conselice},
  {Giavalisco}, {Kassin}, {Lu}, {Lucas}, {Mandelker}, {McIntosh}, {Primack},
  {Ravindranath}, {Barro}, {Ceverino}, {Dekel}, {Faber}, {Fang}, {Koekemoer},
  {Noeske}, {Rafelski}, \& {Straughn}}]{Guo2015}
{Guo}, Y., {Ferguson}, H.~C., {Bell}, E.~F., {et~al.} 2015, \apj, 800, 39

\bibitem[{Gurrola-Ramos {et~al.}(2021)Gurrola-Ramos, Dalmau, \&
  Alarcón}]{gurrola2021unet}
Gurrola-Ramos, J., Dalmau, O., \& Alarcón, T.~E. 2021, IEEE Access, 9, 31742

\bibitem[{{Ivezi{\'c}} {et~al.}(2019){Ivezi{\'c}}, {Kahn}, {Tyson}, {Abel},
  {Acosta}, {Allsman}, {Alonso}, {AlSayyad}, {Anderson}, {Andrew}, {Angel},
  {Angeli}, {Ansari}, {Antilogus}, {Araujo}, {Armstrong}, {Arndt}, {Astier},
  {Aubourg}, {Auza}, {Axelrod}, {Bard}, {Barr}, {Barrau}, {Bartlett}, {Bauer},
  {Bauman}, {Baumont}, {Bechtol}, {Bechtol}, {Becker}, {Becla}, {Beldica},
  {Bellavia}, {Bianco}, {Biswas}, {Blanc}, {Blazek}, {Blandford}, {Bloom},
  {Bogart}, {Bond}, {Booth}, {Borgland}, {Borne}, {Bosch}, {Boutigny},
  {Brackett}, {Bradshaw}, {Brandt}, {Brown}, {Bullock}, {Burchat}, {Burke},
  {Cagnoli}, {Calabrese}, {Callahan}, {Callen}, {Carlin}, {Carlson},
  {Chandrasekharan}, {Charles-Emerson}, {Chesley}, {Cheu}, {Chiang}, {Chiang},
  {Chirino}, {Chow}, {Ciardi}, {Claver}, {Cohen-Tanugi}, {Cockrum}, {Coles},
  {Connolly}, {Cook}, {Cooray}, {Covey}, {Cribbs}, {Cui}, {Cutri}, {Daly},
  {Daniel}, {Daruich}, {Daubard}, {Daues}, {Dawson}, {Delgado}, {Dellapenna},
  {de Peyster}, {de Val-Borro}, {Digel}, {Doherty}, {Dubois},
  {Dubois-Felsmann}, {Durech}, {Economou}, {Eifler}, {Eracleous}, {Emmons},
  {Fausti Neto}, {Ferguson}, {Figueroa}, {Fisher-Levine}, {Focke}, {Foss},
  {Frank}, {Freemon}, {Gangler}, {Gawiser}, {Geary}, {Gee}, {Geha}, {Gessner},
  {Gibson}, {Gilmore}, {Glanzman}, {Glick}, {Goldina}, {Goldstein}, {Goodenow},
  {Graham}, {Gressler}, {Gris}, {Guy}, {Guyonnet}, {Haller}, {Harris},
  {Hascall}, {Haupt}, {Hernandez}, {Herrmann}, {Hileman}, {Hoblitt}, {Hodgson},
  {Hogan}, {Howard}, {Huang}, {Huffer}, {Ingraham}, {Innes}, {Jacoby}, {Jain},
  {Jammes}, {Jee}, {Jenness}, {Jernigan}, {Jevremovi{\'c}}, {Johns}, {Johnson},
  {Johnson}, {Jones}, {Juramy-Gilles}, {Juri{\'c}}, {Kalirai}, {Kallivayalil},
  {Kalmbach}, {Kantor}, {Karst}, {Kasliwal}, {Kelly}, {Kessler}, {Kinnison},
  {Kirkby}, {Knox}, {Kotov}, {Krabbendam}, {Krughoff}, {Kub{\'a}nek},
  {Kuczewski}, {Kulkarni}, {Ku}, {Kurita}, {Lage}, {Lambert}, {Lange},
  {Langton}, {Le Guillou}, {Levine}, {Liang}, {Lim}, {Lintott}, {Long},
  {Lopez}, {Lotz}, {Lupton}, {Lust}, {MacArthur}, {Mahabal}, {Mandelbaum},
  {Markiewicz}, {Marsh}, {Marshall}, {Marshall}, {May}, {McKercher}, {McQueen},
  {Meyers}, {Migliore}, {Miller}, {Mills}, {Miraval}, {Moeyens}, {Moolekamp},
  {Monet}, {Moniez}, {Monkewitz}, {Montgomery}, {Morrison}, {Mueller},
  {Muller}, {Mu{\~n}oz Arancibia}, {Neill}, {Newbry}, {Nief}, {Nomerotski},
  {Nordby}, {O'Connor}, {Oliver}, {Olivier}, {Olsen}, {O'Mullane}, {Ortiz},
  {Osier}, {Owen}, {Pain}, {Palecek}, {Parejko}, {Parsons}, {Pease},
  {Peterson}, {Peterson}, {Petravick}, {Libby Petrick}, {Petry},
  {Pierfederici}, {Pietrowicz}, {Pike}, {Pinto}, {Plante}, {Plate}, {Plutchak},
  {Price}, {Prouza}, {Radeka}, {Rajagopal}, {Rasmussen}, {Regnault}, {Reil},
  {Reiss}, {Reuter}, {Ridgway}, {Riot}, {Ritz}, {Robinson}, {Roby}, {Roodman},
  {Rosing}, {Roucelle}, {Rumore}, {Russo}, {Saha}, {Sassolas}, {Schalk},
  {Schellart}, {Schindler}, {Schmidt}, {Schneider}, {Schneider}, {Schoening},
  {Schumacher}, {Schwamb}, {Sebag}, {Selvy}, {Sembroski}, {Seppala}, {Serio},
  {Serrano}, {Shaw}, {Shipsey}, {Sick}, {Silvestri}, {Slater}, {Smith},
  {Smith}, {Sobhani}, {Soldahl}, {Storrie-Lombardi}, {Stover}, {Strauss},
  {Street}, {Stubbs}, {Sullivan}, {Sweeney}, {Swinbank}, {Szalay}, {Takacs},
  {Tether}, {Thaler}, {Thayer}, {Thomas}, {Thornton}, {Thukral}, {Tice},
  {Trilling}, {Turri}, {Van Berg}, {Vanden Berk}, {Vetter}, {Virieux},
  {Vucina}, {Wahl}, {Walkowicz}, {Walsh}, {Walter}, {Wang}, {Wang}, {Warner},
  {Wiecha}, {Willman}, {Winters}, {Wittman}, {Wolff}, {Wood-Vasey}, {Wu},
  {Xin}, {Yoachim}, \& {Zhan}}]{LSST}
{Ivezi{\'c}}, {\v{Z}}., {Kahn}, S.~M., {Tyson}, J.~A., {et~al.} 2019, \apj,
  873, 111

\bibitem[{Jin {et~al.}(2020)Jin, Meng, Sun, Cui, \& Su}]{raunet}
Jin, Q., Meng, Z., Sun, C., Cui, H., \& Su, R. 2020, Frontiers in
  Bioengineering and Biotechnology, 8

\bibitem[{Kingma \& Ba(2014)}]{kingma2014adam}
Kingma, D.~P. \& Ba, J. 2014 [\eprint[arXiv]{1412.6980}]

\bibitem[{Koekemoer {et~al.}(2011)Koekemoer, Faber, \& Ferguson}]{CANDELS_HST}
Koekemoer, A.~M., Faber, S.~M., \& Ferguson, H.~C. 2011, The Astrophysical
  Journal Supplement Series, 197, 36

\bibitem[{{Laureijs} {et~al.}(2011){Laureijs}, {Amiaux}, {Arduini},
  {Augu{\`e}res}, {Brinchmann}, {Cole}, {Cropper}, {Dabin}, {Duvet}, {Ealet},
  {Garilli}, {Gondoin}, {Guzzo}, {Hoar}, {Hoekstra}, {Holmes}, {Kitching},
  {Maciaszek}, {Mellier}, {Pasian}, {Percival}, {Rhodes}, {Saavedra Criado},
  {Sauvage}, {Scaramella}, {Valenziano}, {Warren}, {Bender}, {Castander},
  {Cimatti}, {Le F{\`e}vre}, {Kurki-Suonio}, {Levi}, {Lilje}, {Meylan},
  {Nichol}, {Pedersen}, {Popa}, {Rebolo Lopez}, {Rix}, {Rottgering},
  {Zeilinger}, {Grupp}, {Hudelot}, {Massey}, {Meneghetti}, {Miller}, {Paltani},
  {Paulin-Henriksson}, {Pires}, {Saxton}, {Schrabback}, {Seidel}, {Walsh},
  {Aghanim}, {Amendola}, {Bartlett}, {Baccigalupi}, {Beaulieu}, {Benabed},
  {Cuby}, {Elbaz}, {Fosalba}, {Gavazzi}, {Helmi}, {Hook}, {Irwin}, {Kneib},
  {Kunz}, {Mannucci}, {Moscardini}, {Tao}, {Teyssier}, {Weller}, {Zamorani},
  {Zapatero Osorio}, {Boulade}, {Foumond}, {Di Giorgio}, {Guttridge}, {James},
  {Kemp}, {Martignac}, {Spencer}, {Walton}, {Bl{\"u}mchen}, {Bonoli},
  {Bortoletto}, {Cerna}, {Corcione}, {Fabron}, {Jahnke}, {Ligori}, {Madrid},
  {Martin}, {Morgante}, {Pamplona}, {Prieto}, {Riva}, {Toledo}, {Trifoglio},
  {Zerbi}, {Abdalla}, {Douspis}, {Grenet}, {Borgani}, {Bouwens}, {Courbin},
  {Delouis}, {Dubath}, {Fontana}, {Frailis}, {Grazian}, {Koppenh{\"o}fer},
  {Mansutti}, {Melchior}, {Mignoli}, {Mohr}, {Neissner}, {Noddle}, {Poncet},
  {Scodeggio}, {Serrano}, {Shane}, {Starck}, {Surace}, {Taylor},
  {Verdoes-Kleijn}, {Vuerli}, {Williams}, {Zacchei}, {Altieri}, {Escudero
  Sanz}, {Kohley}, {Oosterbroek}, {Astier}, {Bacon}, {Bardelli}, {Baugh},
  {Bellagamba}, {Benoist}, {Bianchi}, {Biviano}, {Branchini}, {Carbone},
  {Cardone}, {Clements}, {Colombi}, {Conselice}, {Cresci}, {Deacon}, {Dunlop},
  {Fedeli}, {Fontanot}, {Franzetti}, {Giocoli}, {Garcia-Bellido}, {Gow},
  {Heavens}, {Hewett}, {Heymans}, {Holland}, {Huang}, {Ilbert}, {Joachimi},
  {Jennins}, {Kerins}, {Kiessling}, {Kirk}, {Kotak}, {Krause}, {Lahav}, {van
  Leeuwen}, {Lesgourgues}, {Lombardi}, {Magliocchetti}, {Maguire}, {Majerotto},
  {Maoli}, {Marulli}, {Maurogordato}, {McCracken}, {McLure}, {Melchiorri},
  {Merson}, {Moresco}, {Nonino}, {Norberg}, {Peacock}, {Pello}, {Penny},
  {Pettorino}, {Di Porto}, {Pozzetti}, {Quercellini}, {Radovich}, {Rassat},
  {Roche}, {Ronayette}, {Rossetti}, {Sartoris}, {Schneider}, {Semboloni},
  {Serjeant}, {Simpson}, {Skordis}, {Smadja}, {Smartt}, {Spano}, {Spiro},
  {Sullivan}, {Tilquin}, {Trotta}, {Verde}, {Wang}, {Williger}, {Zhao},
  {Zoubian}, \& {Zucca}}]{Euclid1}
{Laureijs}, R., {Amiaux}, J., {Arduini}, S., {et~al.} 2011, arXiv e-prints,
  arXiv:1110.3193

\bibitem[{Liang {et~al.}(2021)Liang, Cao, Sun, Zhang, Van~Gool, \&
  Timofte}]{swinir}
Liang, J., Cao, J., Sun, G., {et~al.} 2021, in Proceedings of the IEEE/CVF
  international conference on computer vision, 1833--1844

\bibitem[{Liu {et~al.}(2021)Liu, Lin, Cao, Hu, Wei, Zhang, Lin, \&
  Guo}]{swintrans}
Liu, Z., Lin, Y., Cao, Y., {et~al.} 2021, CoRR, abs/2103.14030
  [\eprint{2103.14030}]

\bibitem[{{Lucy}(1974)}]{Lucy1974}
{Lucy}, L.~B. 1974, \aj, 79, 745

\bibitem[{Magain {et~al.}(1998)Magain, Courbin, \& Sohy}]{MCS}
Magain, P., Courbin, F., \& Sohy, S. 1998, The Astrophysical Journal, 494, 472

\bibitem[{Melchior {et~al.}(2018)Melchior, Moolekamp, Jerdee, Armstrong, Sun,
  Bosch, \& Lupton}]{MELCHIOR2018129}
Melchior, P., Moolekamp, F., Jerdee, M., {et~al.} 2018, Astronomy and
  Computing, 24, 129

\bibitem[{Michalewicz {et~al.}(2023)Michalewicz, Millon, Dux, \&
  Courbin}]{STARRED}
Michalewicz, K., Millon, M., Dux, F., \& Courbin, F. 2023, Journal of Open
  Source Software, 8, 5340

\bibitem[{Mohan {et~al.}(2020)Mohan, Kadkhodaie, Simoncelli, \&
  Fernandez-Granda}]{Mohan2020Robust}
Mohan, S., Kadkhodaie, Z., Simoncelli, E.~P., \& Fernandez-Granda, C. 2020, in
  International Conference on Learning Representations

\bibitem[{Nammour {et~al.}(2022)Nammour, Akhaury, Girard, Lanusse, Sureau, Ali,
  \& Starck}]{shapenet}
Nammour, F., Akhaury, U., Girard, J.~N., {et~al.} 2022, A\&A, 663, A69

\bibitem[{nan Xiao {et~al.}(2018)nan Xiao, Lian, Luo, \&
  Li}]{Xiao2018WeightedRF}
nan Xiao, X., Lian, S., Luo, Z., \& Li, S. 2018, 2018 9th International
  Conference on Information Technology in Medicine and Education (ITME), 327

\bibitem[{Ramzi {et~al.}(2023)Ramzi, Michalewicz, Starck, Moreau, \&
  Ciuciu}]{ramzi:hal-03346892}
Ramzi, Z., Michalewicz, K., Starck, J.-L., Moreau, T., \& Ciuciu, P. 2023,
  Journal of Mathematical Imaging and Vision, 65, 240

\bibitem[{{Richardson}(1972)}]{Richardson1972}
{Richardson}, W.~H. 1972, Journal of the Optical Society of America
  (1917-1983), 62, 55

\bibitem[{Ronneberger {et~al.}(2015)Ronneberger, Fischer, \&
  Brox}]{ronnenberger2015Unet}
Ronneberger, O., Fischer, P., \& Brox, T. 2015, CoRR, abs/1505.04597
  [\eprint[arXiv]{1505.04597}]

\bibitem[{{Sattari} {et~al.}(2023){Sattari}, {Mobasher}, {Chartab}, {Kelson},
  {Teplitz}, {Rafelski}, {Grogin}, {Koekemoer}, {Wang}, {Windhorst}, {Alavi},
  {Prichard}, {Sunnquist}, {Gardner}, {Gawiser}, {Hathi}, {Hayes}, {Ji},
  {Mehta}, {Robertson}, {Scarlata}, {Yung}, {Conselice}, {Dai}, {Guo}, {Lucas},
  {Martin}, \& {Ravindranath}}]{Sattari2023}
{Sattari}, Z., {Mobasher}, B., {Chartab}, N., {et~al.} 2023, \apj, 951, 147

\bibitem[{Simard {et~al.}(2002)Simard, Willmer, Vogt, Sarajedini, Phillips,
  Weiner, Koo, Im, Illingworth, \& Faber}]{Simard_2002}
Simard, L., Willmer, C. N.~A., Vogt, N.~P., {et~al.} 2002, The Astrophysical
  Journal Supplement Series, 142, 1

\bibitem[{{Skilling} \& {Bryan}(1984)}]{skilling1984}
{Skilling}, J. \& {Bryan}, R.~K. 1984, \mnras, 211, 111

\bibitem[{Sok {et~al.}(2022)Sok, Muzzin, Jablonka, Marsan, Tan, Alcorn,
  Marchesini, \& Stefanon}]{Sok_2022}
Sok, V., Muzzin, A., Jablonka, P., {et~al.} 2022, The Astrophysical Journal,
  924, 7

\bibitem[{Starck {et~al.}(2015)Starck, Murtagh, \& Bertero}]{Starck2015}
Starck, J.-L., Murtagh, F., \& Bertero, M. 2015, Starlet Transform in
  Astronomical Data Processing, ed. O.~Scherzer (New York, NY: Springer New
  York), 2053--2098

\bibitem[{Starck {et~al.}(1995)Starck, Murtagii, \& Bijaoui}]{MRSupp}
Starck, J.-L., Murtagii, F., \& Bijaoui, A. 1995, Graphical Models and Image
  Processing, 57, 420

\bibitem[{Sureau {et~al.}(2020)Sureau, Lechat, \& Starck}]{sureau2020}
Sureau, F., Lechat, A., \& Starck, J.-L. 2020, \aap, 641, A67

\bibitem[{Tikhonov \& Arsenin(1977)}]{tikhonov1977solutions}
Tikhonov, A.~N. \& Arsenin, V.~Y. 1977, Solutions of ill-posed problems
  (Washington, D.C.: John Wiley \& Sons, New York: V. H. Winston \& Sons),
  xiii+258, translated from the Russian, Preface by translation editor Fritz
  John, Scripta Series in Mathematics

\bibitem[{Vaswani {et~al.}(2017)Vaswani, Shazeer, Parmar, Uszkoreit, Jones,
  Gomez, Kaiser, \& Polosukhin}]{transformer}
Vaswani, A., Shazeer, N., Parmar, N., {et~al.} 2017, in Advances in Neural
  Information Processing Systems, ed. I.~Guyon, U.~V. Luxburg, S.~Bengio,
  H.~Wallach, R.~Fergus, S.~Vishwanathan, \& R.~Garnett, Vol.~30 (Curran
  Associates, Inc.)

\bibitem[{Wang {et~al.}(2004)Wang, Bovik, Sheikh, \& Simoncelli}]{ssim}
Wang, Z., Bovik, A., Sheikh, H., \& Simoncelli, E. 2004, IEEE Transactions on
  Image Processing, 13, 600

\bibitem[{Wang \& Bovik(2009)}]{mse}
Wang, Z. \& Bovik, A.~C. 2009, IEEE Signal Processing Magazine, 26, 98

\bibitem[{Wang {et~al.}(2022)Wang, Cun, Bao, Zhou, Liu, \& Li}]{uformer}
Wang, Z., Cun, X., Bao, J., {et~al.} 2022, in Proceedings of the IEEE/CVF
  Conference on Computer Vision and Pattern Recognition, 17683--17693

\bibitem[{{White, S. D. M.} {et~al.}(2005){White, S. D. M.}, {Clowe, D. I.},
  {Simard, L.}, {Rudnick, G.}, {De Lucia, G.}, {Arag\'on-Salamanca, A.},
  {Bender, R.}, {Best, P.}, {Bremer, M.}, {Charlot, S.}, {Dalcanton, J.},
  {Dantel, M.}, {Desai, V.}, {Fort, B.}, {Halliday, C.}, {Jablonka, P.},
  {Kauffmann, G.}, {Mellier, Y.}, {Milvang-Jensen, B.}, {Pell\'o, R.},
  {Poggianti, B.}, {Poirier, S.}, {Rottgering, H.}, {Saglia, R.}, {Schneider,
  P.}, \& {Zaritsky, D.}}]{white2005}
{White, S. D. M.}, {Clowe, D. I.}, {Simard, L.}, {et~al.} 2005, A\&A, 444, 365

\bibitem[{{Wuyts} {et~al.}(2012){Wuyts}, {F{\"o}rster Schreiber}, {Genzel},
  {Guo}, {Barro}, {Bell}, {Dekel}, {Faber}, {Ferguson}, {Giavalisco}, {Grogin},
  {Hathi}, {Huang}, {Kocevski}, {Koekemoer}, {Koo}, {Lotz}, {Lutz}, {McGrath},
  {Newman}, {Rosario}, {Saintonge}, {Tacconi}, {Weiner}, \& {van der
  Wel}}]{Wuyts2012}
{Wuyts}, S., {F{\"o}rster Schreiber}, N.~M., {Genzel}, R., {et~al.} 2012, \apj,
  753, 114

\bibitem[{Yan {et~al.}(2020)Yan, Zhang, Liu, Zhu, Sun, Shi, \& Zhang}]{deephdr}
Yan, Q., Zhang, L., Liu, Y., {et~al.} 2020, IEEE Transactions on Image
  Processing, 29, 4308

\bibitem[{Yu {et~al.}(2019)Yu, Park, \& Jeong}]{yu2019deep}
Yu, S., Park, B., \& Jeong, J. 2019, in Proceedings of the IEEE Conference on
  Computer Vision and Pattern Recognition Workshops

\bibitem[{Yuan {et~al.}(2021)Yuan, Chen, Wang, Yu, Shi, Jiang, Tay, Feng, \&
  Yan}]{yuan2021tokenstotoken}
Yuan, L., Chen, Y., Wang, T., {et~al.} 2021, Tokens-to-Token ViT: Training
  Vision Transformers from Scratch on ImageNet

\bibitem[{Zamir {et~al.}(2022)Zamir, Arora, Khan, Hayat, Khan, \&
  Yang}]{restormer}
Zamir, S.~W., Arora, A., Khan, S., {et~al.} 2022, in Proceedings of the
  IEEE/CVF Conference on Computer Vision and Pattern Recognition, 5728--5739

\end{thebibliography}

\begin{appendix}

\section{Supplementary figure}
\label{sec:supp_fig}

\begin{figure}[h!]
    \begin{center}
    \includegraphics[width=\columnwidth]{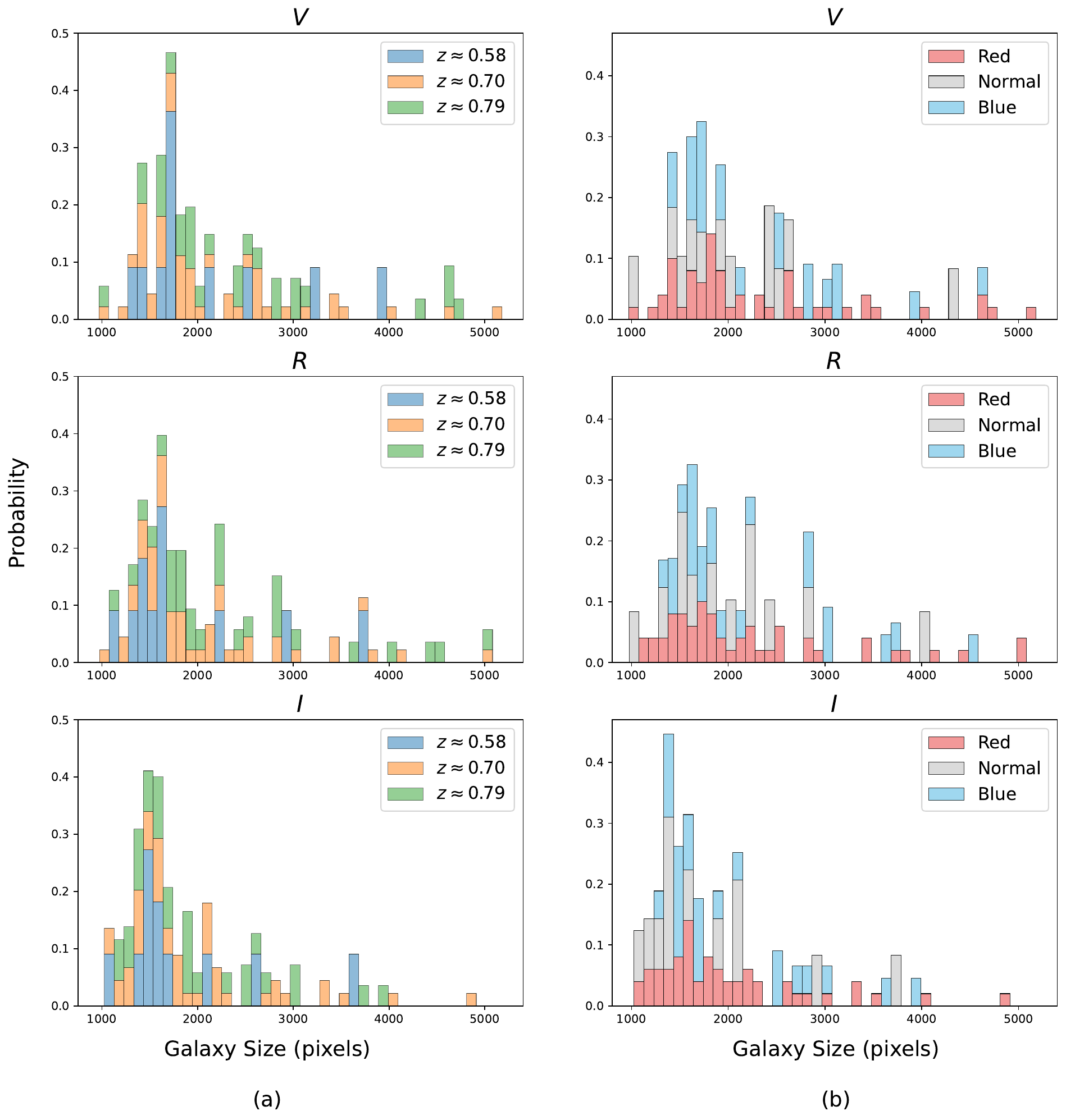}
    \end{center}
    \caption{\label{fig:hist_combined}{Validity check to ensure that the properties of our sample galaxies remain unaffected by size biases. (a): Histogram of galaxy sizes in the $V$-, $R$-, and $I$-bands grouped by their parent cluster redshift: $z\approx0.58$, $z\approx0.70$, $z\approx0.79$. Each coloured bar in the plot represents a specific redshift value, with bars of different redshifts stacked on top of each other. (b): Histogram of galaxy sizes in the $V$-, $R$-, and $I$-bands grouped by their disc colour. Galaxies are classified as `Red' if they are redder, `Normal' if they are comparable, and `Blue' if they are bluer than the field members. Each coloured bar in the plot represents a disc colour category, with bars of different disc colours stacked on top of each other.}}
\end{figure}

As a validity check, we ensured that any insights into the internal properties of our sample galaxies remain unaffected by size biases, whether concerning redshift or disc colours. This is illustrated in Figure \ref{fig:hist_combined}, where (a) represents the histogram of galaxy sizes grouped by their parent cluster redshift and (b) represents the same but grouped by disc colour. Both plots indicate that all galaxies share the same global spatial extent. Each individual legend in the histograms has been normalised such that its probability adds up to one.

\section{Impact of debiasing with multi-resolution support on neural networks}
\label{sec:MRS_appendix}

\begin{figure}[h!]
    \begin{center}
    \includegraphics[width=0.9\columnwidth]{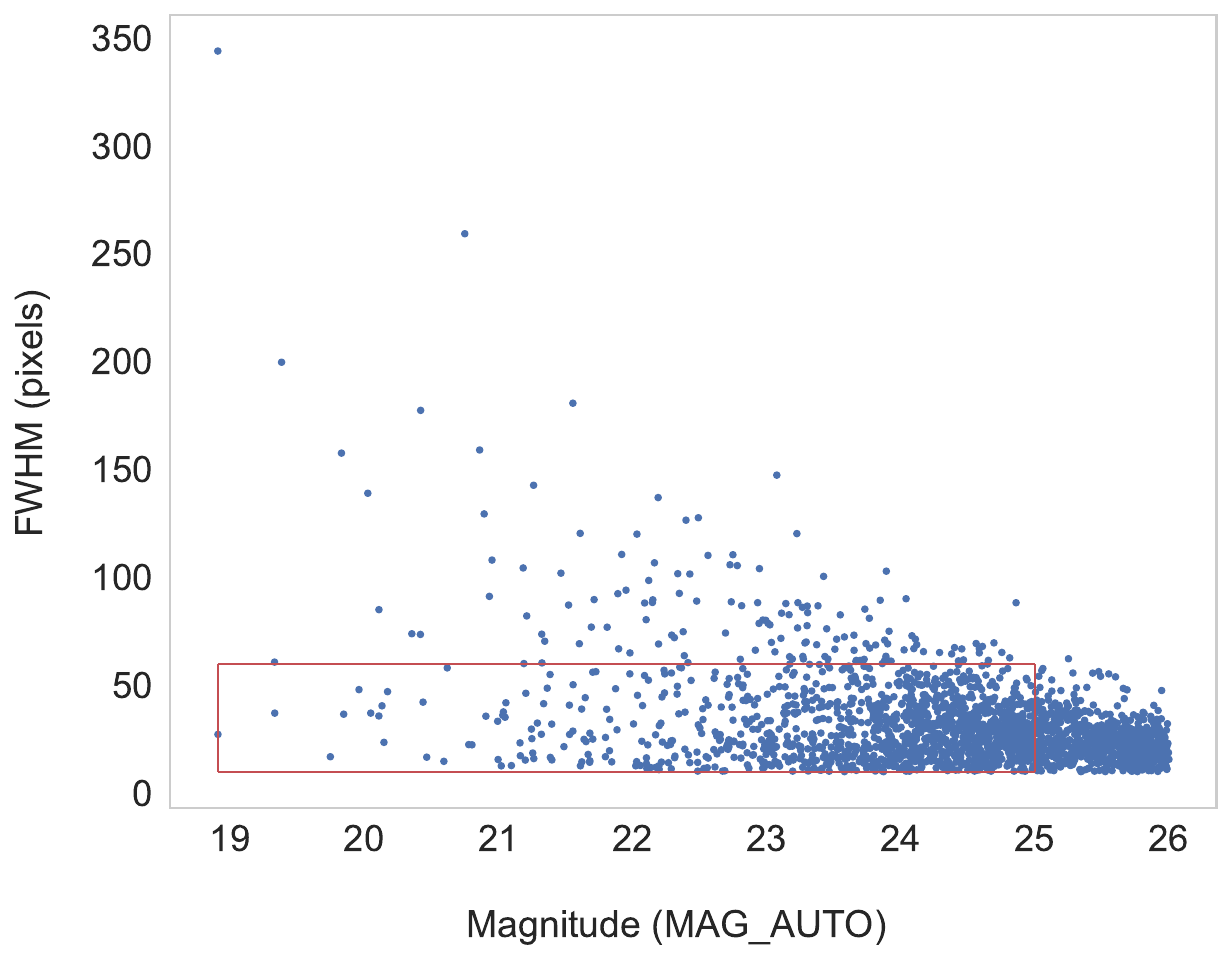}
    \end{center}
    \caption{\label{fig:fwhm_mag}{FWHM vs. magnitude plot for the CANDELS dataset. The red rectangle encloses the $2232$ galaxies selected for the analysis. The limiting magnitude threshold was set at $25$. To eliminate point-sized sources, we applied a minimum FWHM threshold of $10$. A maximum FWHM threshold of $60$ was set to confine the objects within the $128 \times 128$ cutout window.}}
\end{figure}

\begin{figure}[h!]
    \centering
    \subfigure[\makebox{SSIM between the deconvolved outputs and ground truth.}]
    {\label{subfig:sunet1}\includegraphics[width=\linewidth]{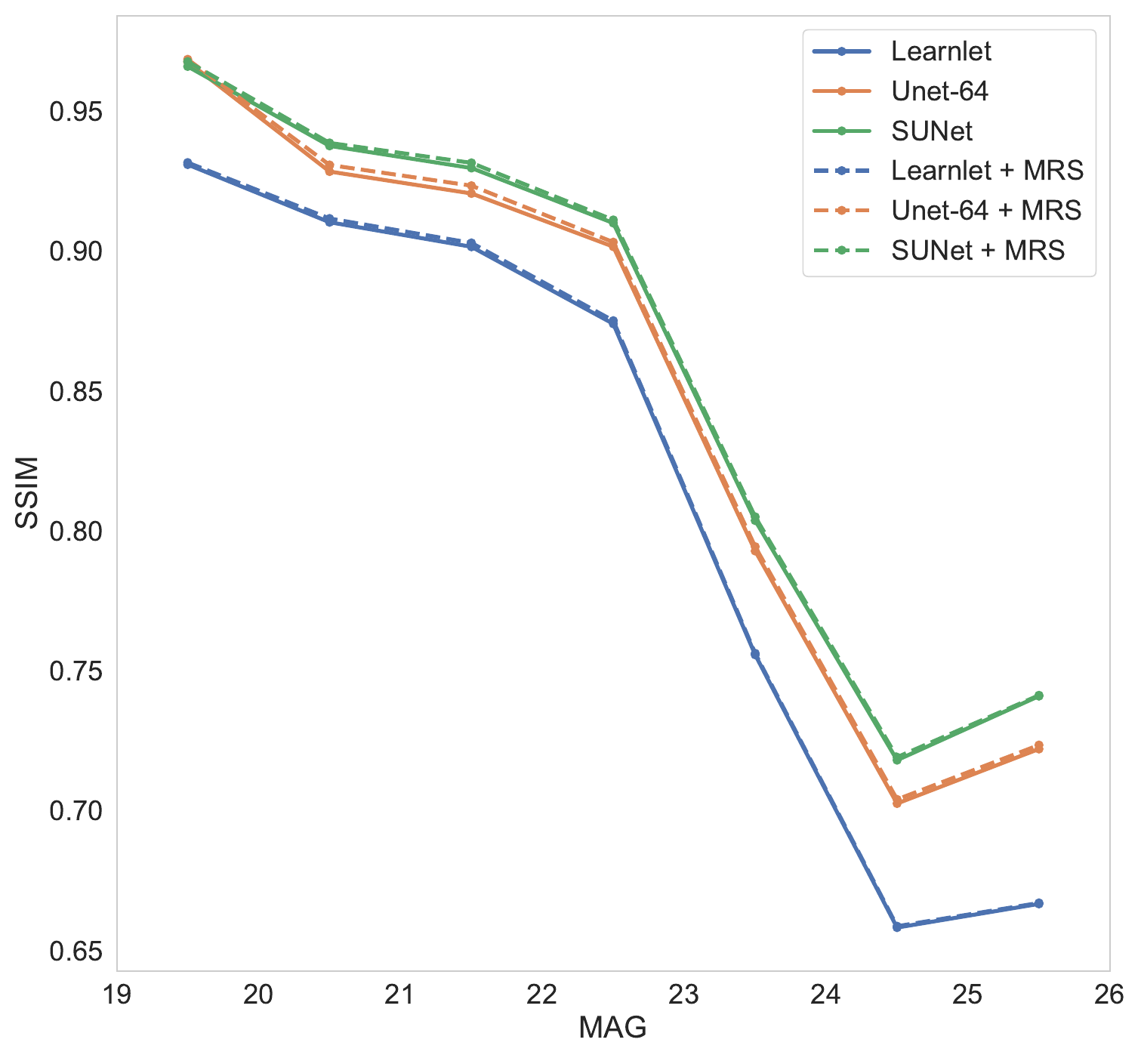}}\\
    
    \subfigure[\makebox{Relative flux error in small-scale structures.}]
    {\label{subfig:sunet2}\includegraphics[width=\linewidth]{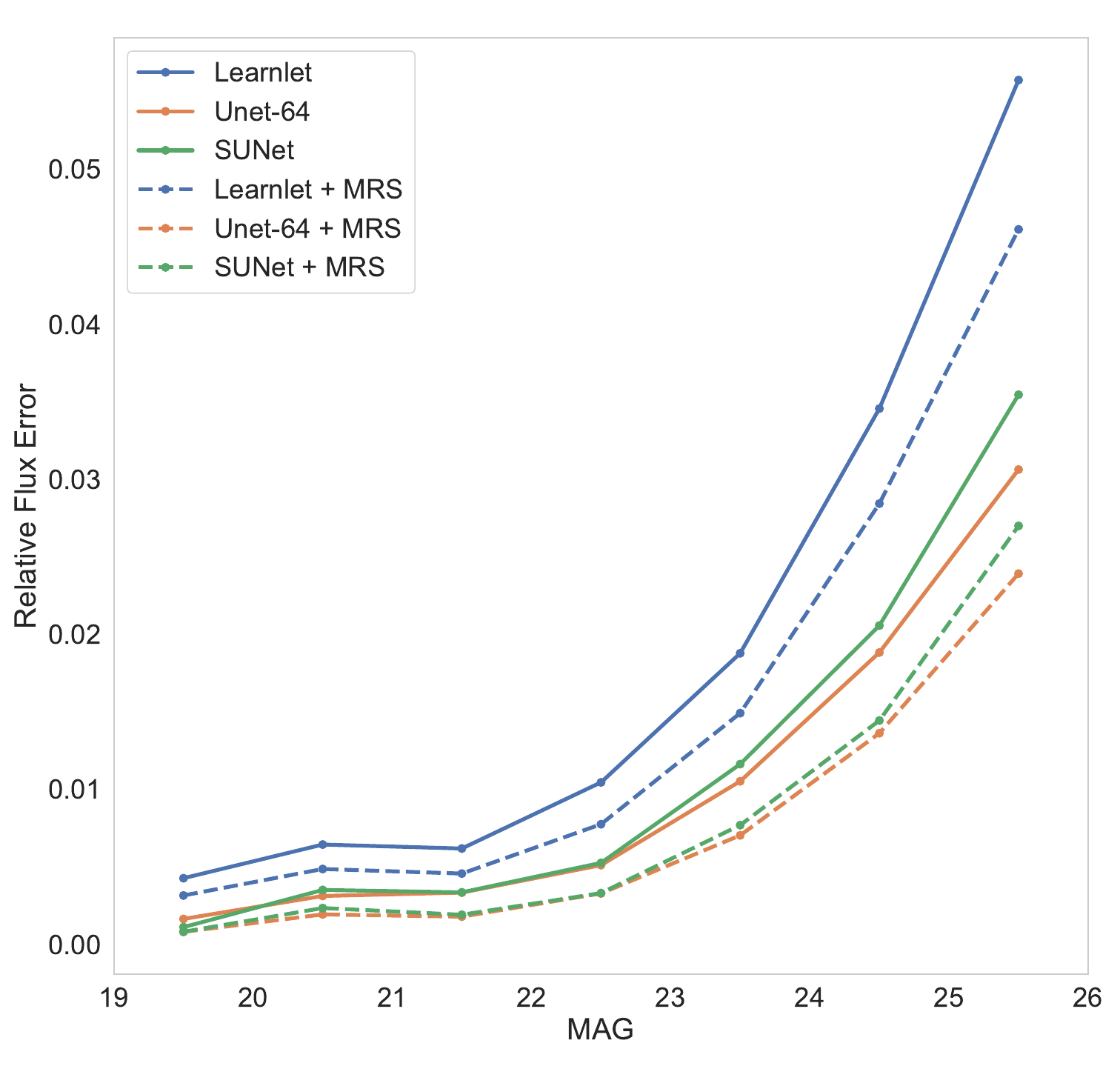}}\\
    
    \caption{Trends in (\ref{subfig:sunet1}) SSIM and (\ref{subfig:sunet2}) flux error as a function of the object magnitude for the three networks: Learnlet, Unet-64, and SUNet. The trends for the original outputs are shown with solid lines, and the trends for the debiased outputs using multi-resolution support are shown with dotted lines. After debiasing, a noticeable enhancement in flux error can be observed for all three networks across a range of magnitudes, with a slight improvement in SSIM.}
    \label{fig:MRS_appendix}
\end{figure}

For a more rigorous study of the impact of debiasing with multi-resolution support on neural networks, we considered three different neural networks with and without the multi-resolution debiasing: Learnlet \citep{ramzi:hal-03346892}, Unet-64 \citep{ronnenberger2015Unet}, and SUNet \citep{sunet}. To provide a quantitative comparison, we computed the SSIM and the flux error in small-scale structures between $2232$ galaxies extracted from CANDELS and their simulated degraded versions, as done in \cite{akhaury2022}. The selection of galaxies based on their FWHM and magnitude is depicted in Figure \ref{fig:fwhm_mag}. To prevent the background noise in the HST images from biasing our metrics, we fit a Gaussian window around each object with an FWHM equal to its catalogue-derived value. The trends in the metrics, depicted in Figure \ref{fig:MRS_appendix} as a function of the object magnitude, indicate an enhancement in SSIM and flux error for all three networks after debiasing.

\section{Hallucinations and the impact of training loss function}
\label{sec:hallucintaions}

\begin{figure}[h!]
    \centering
    \subfigure[\makebox{}]
    {\label{subfig:hall_fwhm}\includegraphics[width=\linewidth]{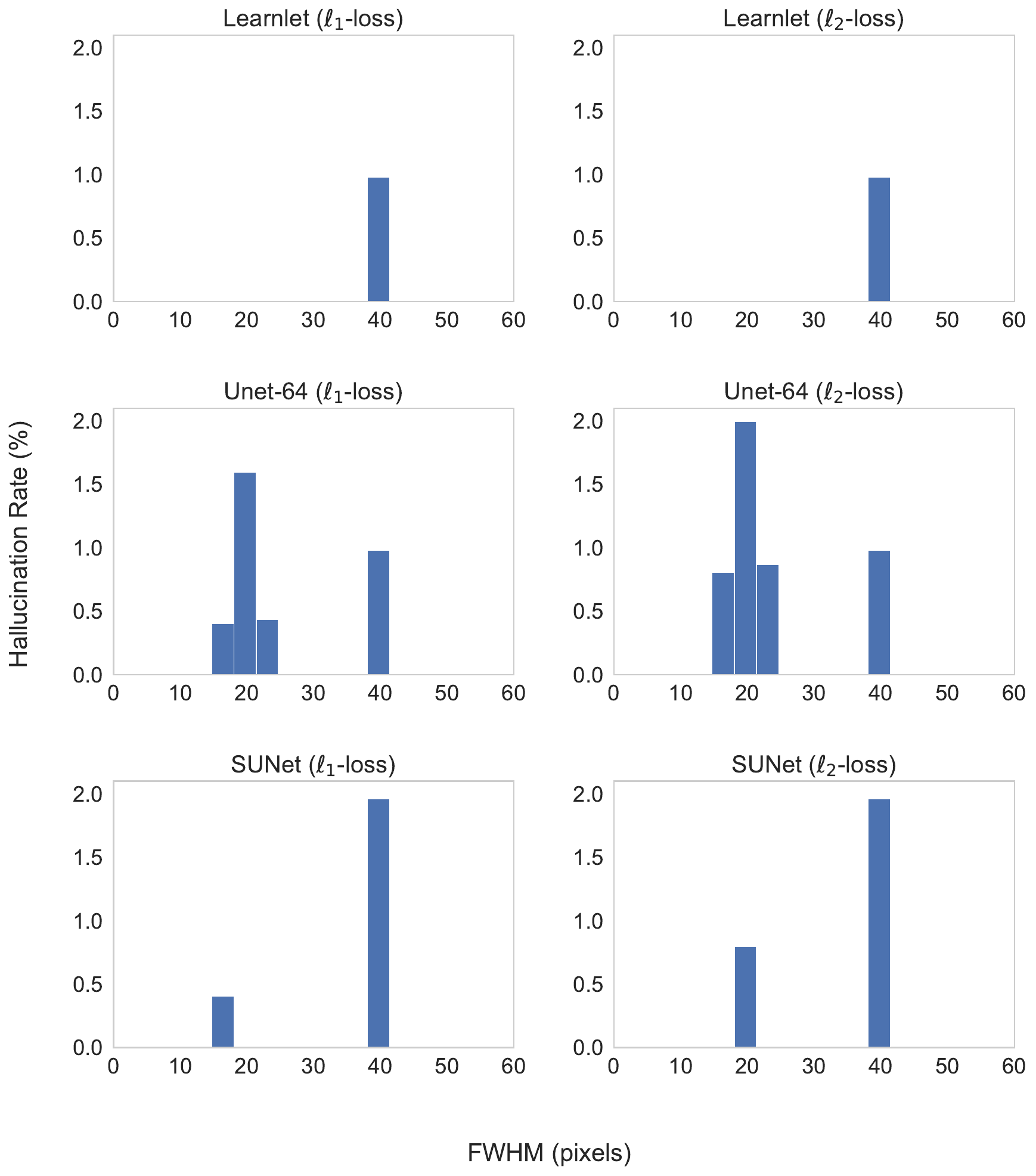}}\\
    
    \subfigure[\makebox{}]
    {\label{subfig:hall_mag}\includegraphics[width=\linewidth]{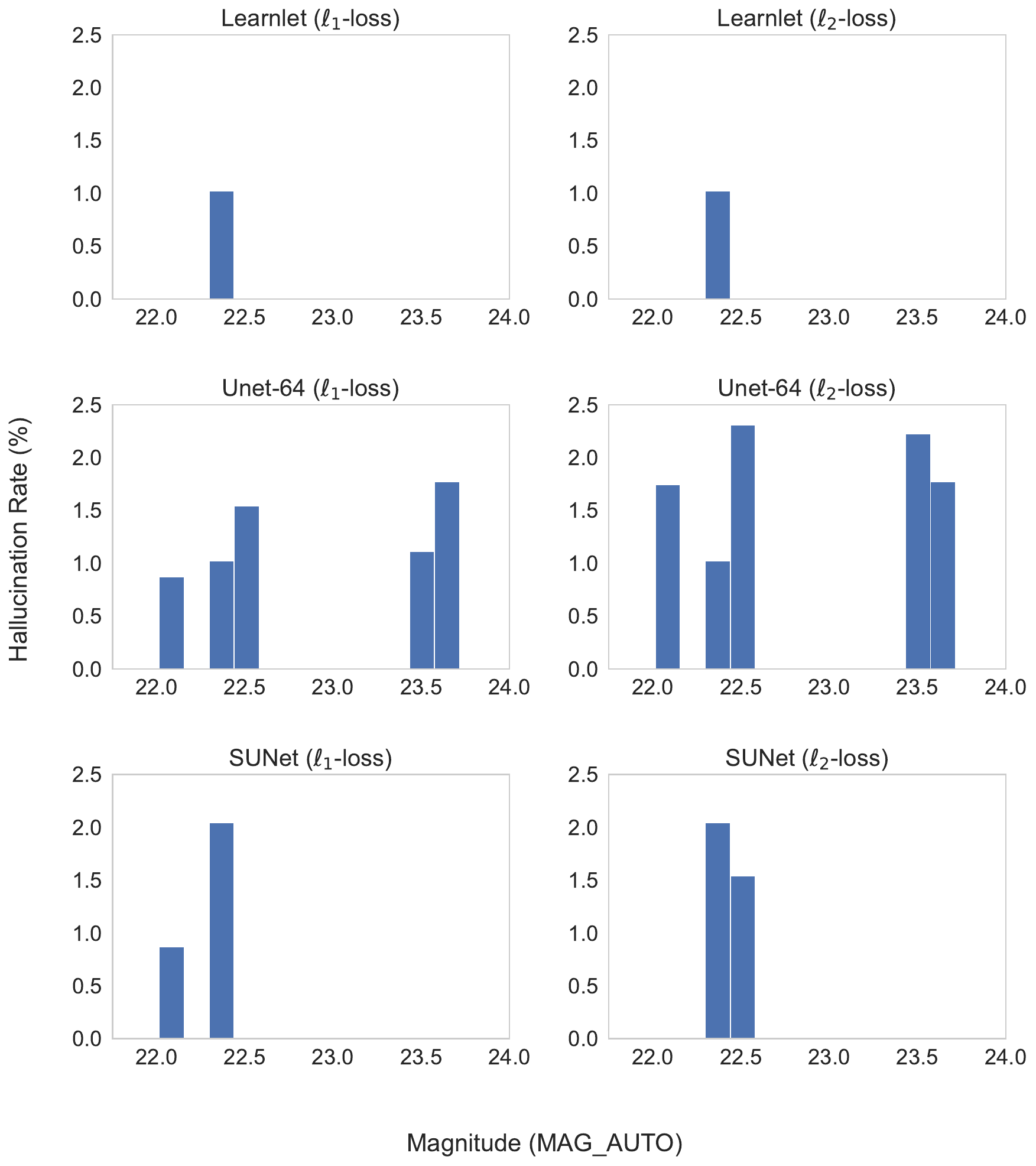}}\\
    
    \caption{Hallucination rate for the three networks—Learnlet, Unet-64, and SUNet—with $\ell_1$ and $\ell_2$ loss as a function of (\ref{subfig:hall_fwhm}) FWHM and (\ref{subfig:hall_mag}) magnitude.}
    \label{fig:hall_appendix}
\end{figure}

To estimate the occurrence of unexpected artefacts or hallucinations introduced by neural networks, we applied the SCARLET detection procedure (as outlined in Section \ref{sec:size_clump}) with a tight $5\sigma$ detection threshold to each scale. To determine the number of false positives, we examined whether the centroid of a detection in the neural network's output fell within a five-pixel radius of the HST detection centroid. Detections failing to meet this criterion were considered false positives. Figure \ref{fig:hall_appendix} illustrates the impact of different loss functions on the hallucination rate as a function of the galaxy FWHM and magnitude for three neural networks: Learnlet, Unet-64, and SUNet. We conducted the experiments on the same test dataset of 2232 galaxies detailed in Appendix \ref{sec:MRS_appendix}. Notably, Unet-64 consistently exhibited improved performance when trained with the $\ell_1$-loss. 

        

\end{appendix}

\end{document}